%% file: main.tex
\documentclass[10pt,conference]{IEEEtran}

\usepackage[normalem]{ulem}
\usepackage{cite}
\usepackage{amsmath,amssymb,amsfonts}
\usepackage{algorithmic}
\usepackage{booktabs}
\usepackage{caption}
\usepackage{listings}
\usepackage[table]{xcolor}
\usepackage{tcolorbox}
\usepackage{textcomp}
\usepackage{boldline}
\usepackage{float}
\usepackage{xspace}
\usepackage{booktabs, makecell, multirow, tabularx}
\usepackage{todonotes} % for comments4
\usepackage{wrapfig,lipsum}
\usepackage{verbatim}
\usepackage{graphicx}
\usepackage{listings}
\usepackage{tikz}
\usepackage{multicol}
\usepackage{wrapfig}
\usepackage{kantlipsum}
\usepackage{boldline}
\usepackage{bbding,captdef}
\usepackage{url}
\newcommand{\ifmap}{\textit{ifmap}\xspace}
\newcommand{\ofmap}{\textit{ofmap}\xspace}
\newcommand{\ifmaps}{\textit{ifmaps}\xspace}
\newcommand{\ofmaps}{\textit{ofmaps}\xspace}

\newcommand{\HuffDuff}{\textit{HuffDuff}\xspace}

\definecolor{dkgreen}{rgb}{0,0.6,0}
\definecolor{gray}{rgb}{0.5,0.5,0.5}
\definecolor{mauve}{rgb}{0.58,0,0.82}

\newcommand*\circled[1]{\tikz[baseline=(char.base)]{
            \node[shape=circle,fill,inner sep=0.3pt] (char) {\textcolor{white}{#1}};}}

\newcommand*\circlenew[1]{\tikz[baseline=(char.base)]{
            \node[shape=circle,draw,inner sep=0.3pt] (char) {#1};}}      

\usepackage{tikz}
\newcommand*\emptycirc[1][1ex]{\tikz\draw (0,0) circle (#1);} 

\newcommand*\fullcirc[1][1ex]{\tikz\fill (0,0) circle (#1);}

\newcommand{\name}{{\em SparseLock}\xspace}
\newcommand{\fname}{{\em SparseLock}\xspace}
\def\BibTeX{{\rm B\kern-.05em{\sc i\kern-.025em b}\kern-.08em
    T\kern-.1667em\lower.7ex\hbox{E}\kern-.125emX}}

\begin{document}

\title{SparseLock: Securing Neural Network Models in Deep Learning Accelerators}

\author{\IEEEauthorblockN{Nivedita Shrivastava}
\IEEEauthorblockA{Electrical Engineering Department\\
Indian Institute of Technology, Delhi, India\\
Email: nivedita.shrivastava@ee.iitd.ac.in}
\and
\IEEEauthorblockN{Smruti Ranjan Sarangi}
\IEEEauthorblockA{Electrical Engineering Department\\
Indian Institute of Technology, Delhi, India\\
Email: srsarangi@cse.iitd.ac.in}
}

\date{}
\maketitle

\thispagestyle{empty}

\input{abstract}
\input{intro.tex}
\input{background}

\input{table}
\input{threatModel}
\input{attack}

\input{characterization_attack}
\input{characterization}
\input{design}

\input{evaluation}
\input{securityEvaluation}
\input{relatedWork.tex}

\input{conclusion}

{\footnotesize
\bibliographystyle{IEEEtran}
\bibliography{ref}
}

\end{document}

%% file: abstract.tex
\begin{abstract}
    Securing neural networks (NNs) against model extraction and parameter exfiltration attacks is an important problem
primarily because modern NNs take a lot of time and resources to build and train. We observe that there are no
countermeasures (CMs) against recently proposed attacks on sparse NNs and there is no single CM that effectively
protects against all types of known attacks for both sparse as well as dense NNs. In this paper, we propose \fname, a
comprehensive CM that protects against all types of attacks including some of the very recently proposed ones for which
no CM exists as of today. We rely on a novel compression algorithm and binning strategy. Our security guarantees are
based on the inherent hardness of bin packing and inverse bin packing problems. We also perform a battery of statistical
and information theory based tests to successfully show that we leak very little information and  side channels in our
architecture are akin to random sources. In addition, we show a performance benefit of 47.13\% over the nearest
competing
secure architecture.
\end{abstract}

%% file: intro.tex
\section{Introduction}
Deep learning (DL) algorithms are all-pervading as of today and are considered the de facto methods of choice for
image classification~\cite{classification}, advanced security systems~\cite{security},
autonomous driving technology~\cite{automobile,UAV} and  speech recognition~\cite{speech}.
The DL models embody a significant amount of intellectual property (IP) because it takes a lot of money
and effort to collect data, design and train the models~\cite{securator,scedl}.  Hence, it is very important to protect them
against attackers, who have a significant incentive in stealing the architectural parameters and weights of the models.
Furthermore, having some information about the DL model greatly increases the likelihood of designing successful
adversarial attacks~\cite{adversarial}, membership
inference attacks~\cite{membership} and bit-flip attacks~\cite{deephammer}. 

We examined all attacks on neural networks targeting model theft (architecture and values), which have been proposed over
the last decade. They can be classified into three classes: model stealing side-channel attacks (SCAs) for dense NNs
(neural networks)
({\em Type A}), model stealing SCAs for sparse NNs ({\em Type B}), and data theft attacks ({\em Type C}).  Reverse
engg.~\cite{reverse} and CacheTelepathy~\cite{cachetelepathy} propose stealing the dense NN model architecture for
accelerators and CPUs using bus-snooping and cache-based SCAs, respectively. They rely on a correlation between
leaked memory access
patterns and the dimensions of the NN. Other
proposals~\cite{deepsniffer,neurounlock} use various GPU-based side-channels, such as GPU context switches and bus snooping,
to steal memory access patterns. These {\em Type A} attacks ultimately rely on the correlation between the NN's architecture
and the information gathered from side channels of various kinds.
However, Dingqing et al.~\cite{huffduff} highlight that these attacks do not work for sparse NNs or works that 
involve model compression. They have thus recently proposed a
new attack called {\em HuffDuff}, which works with sparse NNs and compressed models as well. We classify this type of an attack as a
{\em Type B} attack. Additionally, an adversary may also attempt to tamper with or steal the model weights.
For instance, the
recently proposed 
DeepFreeze~\cite{deepfreeze} retrieves model parameters using cold-boot attacks. Similarly, DeepSteal~\cite{deepsteal} attack uses the rowhammer attack to design an efficient method for stealing weights of a neural network. Such attacks targeting the values of
the model, and are thus classified as {\em Type C} attacks. Table~\ref{tab:survey} shows a comparative summary of some of the 
recent work in this space.

\begin{table*}[!htb]
    \centering
    \footnotesize
    \begin{tabular}{|l|l|l|l|l|l|l|}
    \hline
    \rowcolor{blue!10}
      \textbf{Countermeasure}  & \textbf{Venue/Year} & \textbf{Comp-} & \textbf{Type A} & \textbf{Type B} & \textbf{Type C} & \textbf{Information theoretic} \\
      \cline{4-5}
      \rowcolor{blue!10}
    &  & \textbf{ression} & {\em Dense} & {\em Sparse}  & \textbf{(Encryption)} & \textbf{analysis}\\
       \hline      
         {\em NPUFort\cite{npufort}} & CF'19 &                     $\times$  & $\times$ & $\times$ & Partial & $\times$ \\
         {\em Sealing\cite{sealing}} & DAC'21 &                    $\times$  & $\times$ & $\times$ & Partial & $\times$ \\
           {\em NeurObfuscator}\cite{neurobfuscator} & HOST'21  &  $\times$ & \checkmark & $\times$ & $\times$  & $\times$\\
         {\em DNNCloak}~\cite{dnncloak} & ICCD'22               &   \checkmark &   \checkmark & $\times$ & Partial & $\times$\\
        {\em ObfuNAS\cite{obfunas}} & ICCAD'22 &                   $\times$  &  \checkmark & $\times$ & $\times$ & $\times$ \\
         {\em TNPU}~\cite{tnpu} & HPCA'22 &                        $\times$ & $\times$ & $\times$ & Full & $\times$\\
        {\em Securator}~\cite{securator} & HPCA'23 &               $\times$ & \checkmark & $\times$ & Full & $\times$\\
           \fname & - &                                            \checkmark & \checkmark & \checkmark & Full & \checkmark\\ 
         \hline
    \end{tabular}
    \caption{A survey of countermeasures for different types of attacks.  (SCAs $\rightarrow$ Side-Channel Attacks)}
    \label{tab:survey}
\end{table*}

{\em Type A} attacks~\cite{reverse,deepsniffer} rely on subtle architectural side channels that one would normally ignore.
Such attacks accurately remove the noise and figure out the parameters of the network with relatively few
attempts. They are effective and work quite surreptitiously. {\em Type B} attacks for sparse NNs use similar 
techniques, albeit with different side channels.
Sometimes, they apply
specially crafted inputs to the neural network such that specific features such as the number of non-zero elements
are exfiltrated via the side
channels. {\em Type C} attacks are easier to address by themselves 
because often memory encryption with freshness guarantees is all
that is needed.

Countermeasures (CMs) have failed to effectively address the problem. Some popular
approaches~\cite{neurobfuscator,dnncloak} propose methods that lead to much more work such as splicing and combining layers,
increasing the size of all the layers and adding random noise (additional dummy computations). Many of these approaches
fail to work with the recently proposed attacks because they effectively filter out the additional zero-valued elements
that these approaches add. NPUFort~\cite{npufort}, TNPU~\cite{tnpu}, MGX~\cite{mgx} and Securator~\cite{securator} rely on the
encryption of weights; however, they don't plug such architectural side channels that primarily rely on data-level
discontinuities ($1 \rightarrow 0$ and $0 \rightarrow 1$ transitions).  Moreover, we could not find any
formal proofs or rigorous empirical analyses in prior work that quantify the degree of security that is
provided. We argue that without some formal reasoning or information theory based analysis, we might erroneously
overestimate the degree of security provided by popular countermeasures (based on our experiments).
{\bf Question: Can we provide a countermeasure (CM) that targets all three kinds of attacks and
can we prove using rigorous information theoretic analysis that our CM is indeed secure?}

We claim that our proposed technique \fname works for all kinds of attacks. Furthermore,
to the best of our knowledge, there is no attack on NN architectures that is  not prevented by our method, subject
to reasonable assumptions (elaborated in Section~\ref{sec:threat}). 
The insight behind our work is as follows. We observe that compression can serve as a method of obfuscation and
can be a potential source of randomness. Needless to say, it still leaks a great deal of information and this has
been exploited by attacks notably \HuffDuff~\cite{huffduff}. However, we augment this basic primitive by grouping together
a set of compressed (and then encrypted) tiles (blocks of bytes) into bins, where a {\em bin} is an atomic unit
in our scheme (its size is fixed). 
Given that the contents of a bin are fetched together, and its contents are encrypted, the scheme is secure.
It is not possible to extract any information easily; we implicitly rely on the hardness on the inverse bin-packing problem~\cite{IBPP}.
Additionally, we show that we leak very little amount of information empirically, and our side channels behave more or less as random sources 
(exact details in Section~\ref{sec:security}).

Our technique relies on a new compression algorithm primarily tailored for NN data.
The first compression level tries to bring out the key features in the data by encoding the data such that the second
level can easily compress it.  We compare our compression algorithms with most of the popular HW-based compression
algorithms proposed in the literature in Section~\ref{sec:char}, and show that our compression  ratios are $2 \times$
better.  Even though the primary motive of our scheme was to address the recently proposed attacks, and use an efficient
compression algorithm as the source of randomness, we also ended up increasing the performance by $1.71\times$ of our
DNN (deep NN) hardware as compared to the state of the art mainly because our compression algorithm is more efficient and  the
critical path is shorter due to enhanced hardware-level parallelism. 

To the best of our knowledge, there is no other competing work that guarantees all of the following: encryption of all
weights and model parameters, side-channel leakage protection (all types of attacks: A, B and C), 
model compression, and a thorough information
theory-based security analysis. 

The specific \textbf{list of contributions in this paper} is as follows: 
\circlenew{1} A comprehensive experimental
study of memory-based side-channel attacks (SCAs) targeting sparse as well as dense accelerators. 
\circlenew{2} An extensive
characterization of state-of-the-art compression algorithms, followed by the design and implementation of a 
compression scheme based on the insights gained from the aforementioned analysis. 
\circlenew{3} A novel approach to
enhance the security of both sparse and dense accelerators. The architectural hints are obfuscated using a combination
of tiling, compression-then-encryption, and a novel binning-based approach. 
\circlenew{4} A detailed performance analysis of
\fname, which shows a $47.31\%$ improvement over the nearest competing work (DNNCloak). 
\circlenew{5} An extensive analysis based on
statistical randomness tests and information theory-based analysis for both the sparse and dense accelerators.

$\S$\ref{sec:back} provides the necessary background. $\S$\ref{sec:threat} outlines the
threat model for our scheme. $\S$\ref{sec:huffduff} provides the details of the side-channel based model architecture
stealing attacks for sparse and dense NNs. $\S$\ref{sec:char} provides a characterization of compression schemes. $\S$\ref{sec:hw}
presents the proposed design, $\S$\ref{sec:eval} presents the performance results, $\S$\ref{sec:security} presents a
detailed security analysis and $\S$\ref{sec:RW} presents the related work. We finally conclude in
$\S$\ref{sec:conclusion}. 

%% file: background.tex
\section{Background} \label{sec:back}

\subsection{Convolution Neural Networks (CNNs)}
The convolution operation serves as the core component of a CNN (most popular NNs, as of today). 
To execute the convolution process, we sequentially
traverse the input and output feature maps (referred to as an \ifmap and \ofmap, respectively), and apply the convolution
operation. The \ofmap matrix \textit{O} is the result of a convolution operation between a filter matrix {\em F} of size ($R \times S$) and a 2D \ifmap matrix $I$ with a single channel. The value of a single element of {\em O} is computed as -

%\fixme{This equation is wrong. What is C? ceiling-floor. How can you have small x and capital X. Very confusing. Use the
%same terminology as the adv. arch. book.}
\begin{equation}
    \textit{O}[h][w] = \sum_{r=0}^{R-1} \sum_{s=0}^{S-1} \textit{F}[r][s] \times \textit{I}[h+r][w+s]
\end{equation}

Due to the limited capacity of on-chip buffers, elements in an \ifmap, an \ofmap, and a filter are often grouped into
{\em tiles}~\cite{timeloop}. Each tile is granted a unique identifier known as the ``{\em Tile ID}''. 

%REFERENCE: https://ieeexplore.ieee.org/stamp/stamp.jsp?tp=&arnumber=9643728

\subsection{Compression Techniques}
Compression techniques can be categorized into three distinct groups: generic compression, hardware-based memory
compression and compression approaches specifically designed for neural networks.  Generic compression algorithms such
as {\em Huffman coding}~\cite{huffman}, {\em zstd}~\cite{zstd}, and {\em zlib}~\cite{zlib}, 
utilize methods like dictionary coding and entropy
coding. Ko et al.~\cite{lane} highlight that these schemes are not suitable for large NN models as
they are too slow. 

Hardware-friendly compression methods such as Cache Packer~\cite{cpack} ({\em C-Pack}), Frequent-pattern
compression~\cite{fpc} ({\em FPC}), and Base-delta-immediate~\cite{bdi} ({\em BDI}) are
comparatively much faster. Table ~\ref{tab:compressAlgo} shows a short summary.

\begin{table}[!htb]
    \centering
    \footnotesize
    \begin{tabular}{|p{1.4cm}|p{6.2cm}|}
    \hline
    \rowcolor{blue!10}
    \textbf{Algorithm} & \textbf{Description} \\
    \hline
    \multicolumn{2}{|c|}{\textbf{General compression}} \\
    \hline
        Huffman~\cite{huffman} &  Translates the data into binary
codes. The code for the character with the highest frequency is the shortest \\
        RLE~\cite{rle} & Run-length encoding\\
        zstd~\cite{zstd} & LZ77, Huffman or run-length encoding\\
\hline
    \multicolumn{2}{|c|}{\textbf{Hardware compression}} \\
    \hline
        FPC~\cite{fpc} & Frequently occurring patterns are assigned a specific code. \\
        BDI~\cite{bdi} & Encodes a block of $n$ values using a base value and $(n-1)$ deltas\\
        C-Pack~\cite{cpack} & Static compact coding for frequently appearing data and dictionary-based compression for the rest\\
         \hline
    \multicolumn{2}{|c|}{\textbf{Neural network compression}}\\
    \hline
         Pruning & Removing unnecessary weights from an NN model\\
         Quantization & Reducing the precision of weights/biases\\
     \hline    
    \end{tabular}
    \caption{Summary of various lossless compression schemes}
    \label{tab:compressAlgo}
\end{table}

\subsubsection*{\underline{Compression of NN Models}}

Modern NN compression techniques typically mix pruning and quantization. {\em Pruning} is the process of
removing irrelevant weights from the NN model. It increases the sparseness (number of zeros) in the model
(refer to Table~\ref{tab:sparse}).

\begin{wraptable}[9]{r}{0.5\linewidth}
   \footnotesize
    \begin{tabular}{|l|p{0.1\textwidth}|}
    \hline
    \rowcolor{blue!10}
    \textbf{Encoding} & \textbf{Accelerator}\\
   \hline
      CSC   &  Eyeriss\cite{eyerissv2}\\
      CSR  &  SCNN\cite{scnn}, FSA\cite{fsa} \\
      Bit-mask & SparTen\cite{sparten} \\
      Offset-based & Swallow\cite{swallow} \\
      \hline
    \end{tabular}
    \caption{Popular sparse encodings used in accelerators}
    \label{tab:sparse}
\end{wraptable}

Pruned models yield sparse \ofmaps and \ifmaps, which can be represented easily using several compressed
formats:
 bit-masks\cite{sparten}, CSC (compressed sparse column)\cite{eyeriss} and CSR (compressed sparse 
row)\cite{scnn}.

\subsection{Fisher Information (FI)}

Information-theoretic metrics (such as the Fisher and mutual information) have been extensively used in the
traditional cryptography literature to
quantify information
leaks~\cite{fisher1,fisher2,fisher3,fisher4,fisher5,mayer2006fisher,MIleakage,MIleakage2,MI3,MI4,MI5}. FI is  a
classical
metric, which measures the amount of information that an observable random variable $X$ carries about an
unknown parameter $\theta$ (e.g., parameters of a layer). Let $X$ be a random variable and $x$ be a realization of the random variable $X$. 

The FI is the variance of the $\frac{\partial }{\partial \theta} {ln (f(X \mid \theta))}$, operator E represents the expectation, and $f(X \mid \theta)$ is the likelihood function. We evaluate this at a fixed parameter value of $\theta$ --

%\fixme{This formula is wrong. Have the right formula. Don't confuse between conditional and a joint distribution.}

\begin{equation}
    FI(\theta) = E{\left(\frac{\partial }{\partial \theta} {ln (f(X \mid \theta))}\mid\theta \right)}^2
\end{equation}

%If $p(x \mid \theta)$ is the same for all
%$\theta$, then $x$ does not provide any information about $\theta$. If $p(x \mid \theta)$ is non-zero only for a single
%$\theta$, then, $x$ provides all the necessary information about $\theta$.
The above equation for discrete data is as follows. 
\begin{equation}
    FI(\theta) =   \sum_{i=1}^{n}{\frac{{{(p_{i+1}-p_i)}}^2}{p_i} }
\end{equation}

Here, $f(X=x_i \mid \theta) = p_{i}$, for a given $\theta$. 
Suppose an adversary utilizes an unbiased
estimator to predict the parameter $\theta$. The Cramer–Rao bound states that the inverse of the Fisher information is a lower bound on the variance (Var) of any unbiased estimator of $\theta$. 
For any unbiased estimator $T$ of $\theta$ -
\begin{equation}
    Var(T) \geq \frac{1}{FI(\theta)}
\end{equation}

\subsection{The Boundary Effect in CNNs}
The boundary effect occurs mainly due to  the image border. We can represent the boundary effect using a simple
mathematical equation.  Let $I \in \mathbb{R}^n$ represent a 1-D image with $n$ elements and a single channel. Let $F
\in \mathbb{R}^{2k+1}$ be a filter with $(2k+1)$ elements and a single channel.  We can express the output of a basic 1D
convolution operation using Equation \ref{equ:conv1D}

\begin{equation}\label{equ:conv1D}
    O[t] = \sum_{j=-k}^{k}{F[j] \times I[t-j]}
\end{equation}

Let us assume that the convolution operator slides only along the center of the filter on all existing image values.
We observe that except for the boundary positions, the number of summations in Equation \ref{equ:conv1D} will remain
$2k+1$. The count of sums {\em `C'} in an \ofmap can be represented using Equation \ref{equ:boundary}~\cite{kayhan2020translation}.

\begin{equation} \label{equ:boundary}
  C[a] =
    \begin{cases}
      a+k & \text{if $a \in [1,k]$}\\
      n-a+(k+1) & \text{if $a \in [n-k ,n]$}\\
      2k+1 & \text{otherwise}
    \end{cases}       
\end{equation}
Here, {\em n} is the number of elements in the \ifmap, and {\em 2k+1} is the number of elements in the filter. 
%\fixme{Explain the terms in this equation.}

%% file: table.tex
\begin{table*}[ht]
\centering
\footnotesize
\begin{tabular}{ccccccc}

0$\%$ sparsity & 0$\%$ sparsity & 0$\%$ sparsity & 30$\%$ sparsity & 30$\%$ sparsity &  30$\%$ sparsity & 60$\%$ sparsity
\\
($3$ elements) & ($5$ elements) & ($7$ elements) & ($3$ elements) & ($5$ elements) &  ($7$ elements) & ($7$ elements)
\\
\hline
\includegraphics[scale=0.1]{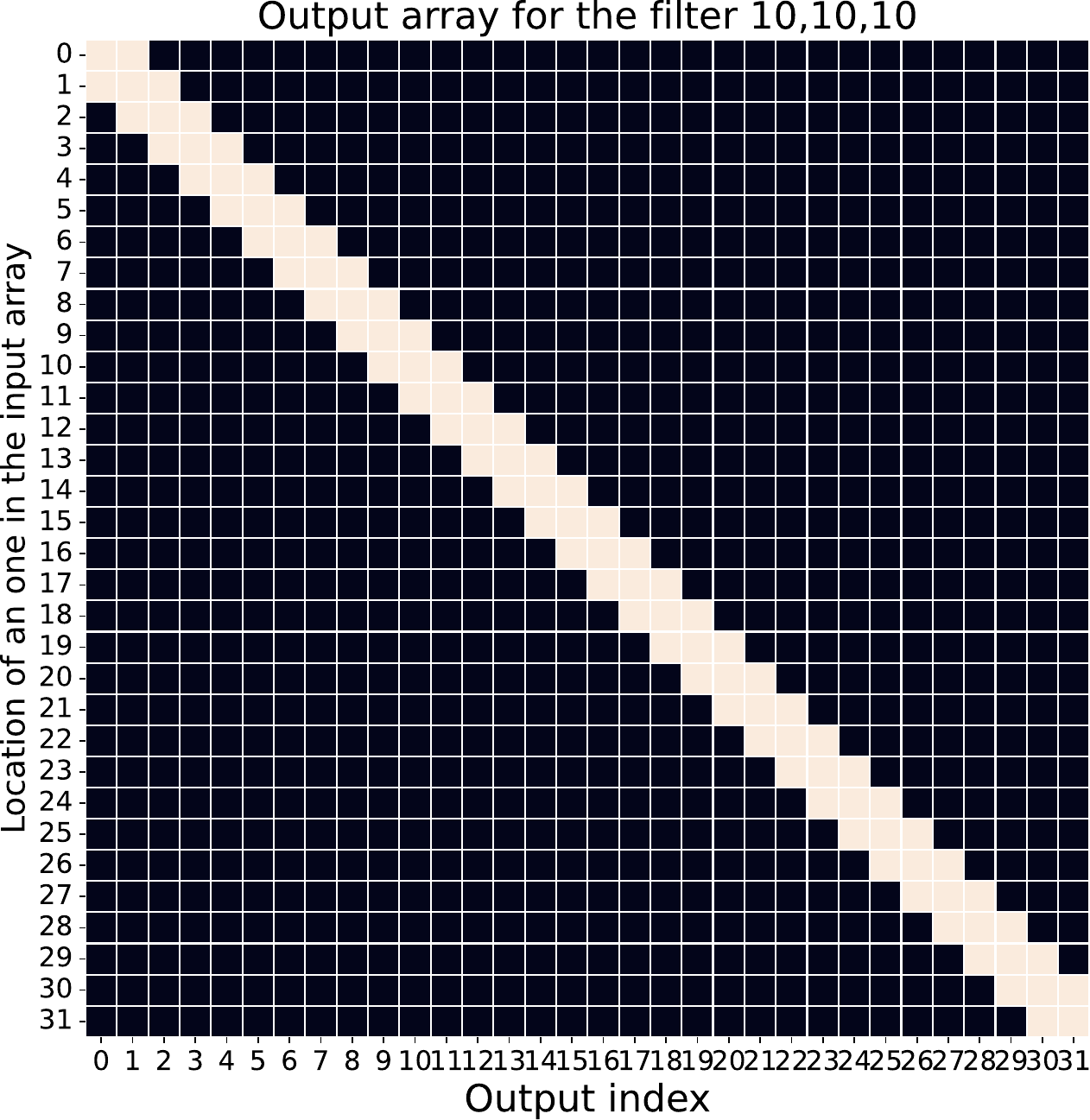} &\includegraphics[scale=0.1]{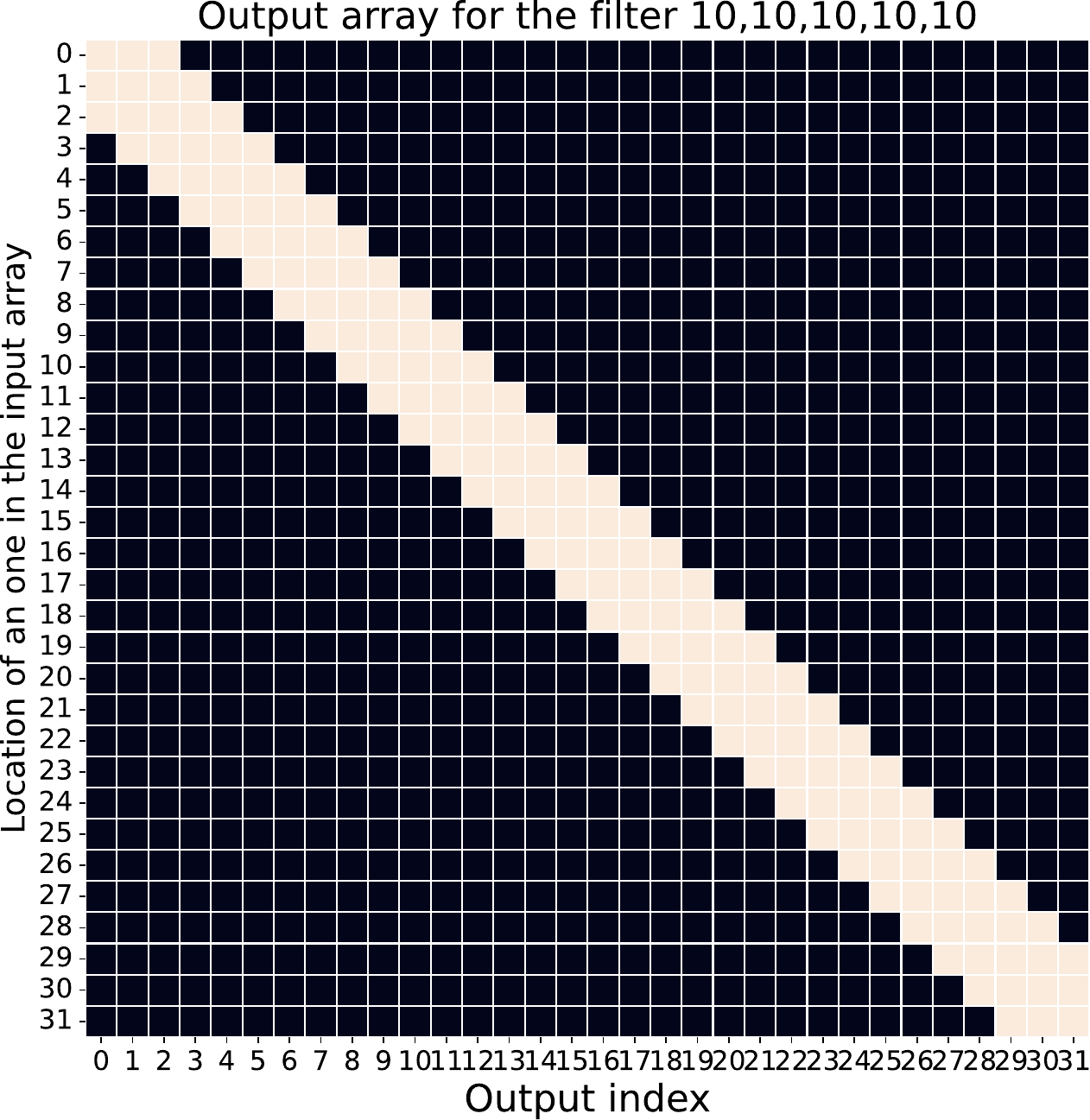}&\includegraphics[scale=0.1]{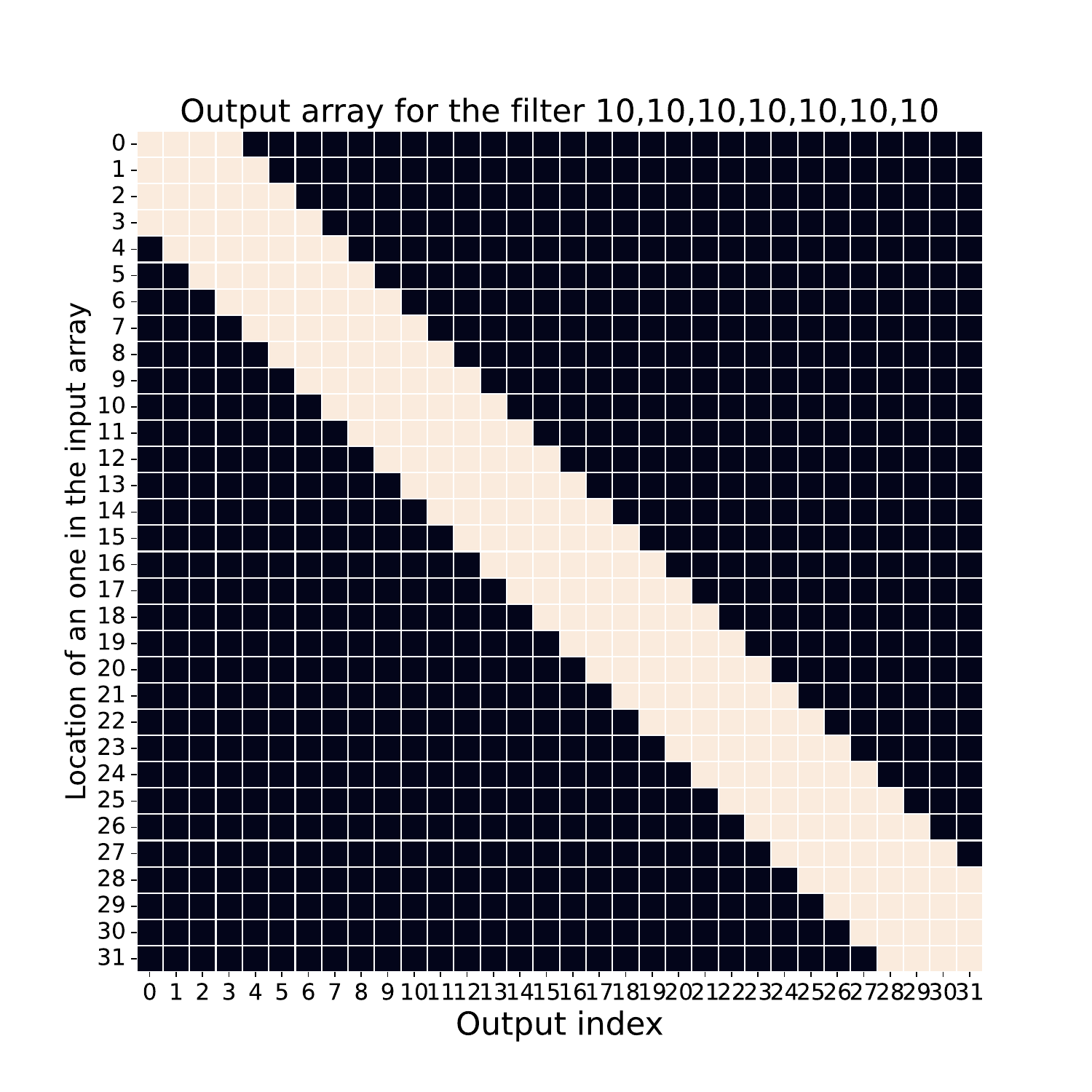}&\includegraphics[scale=0.1]{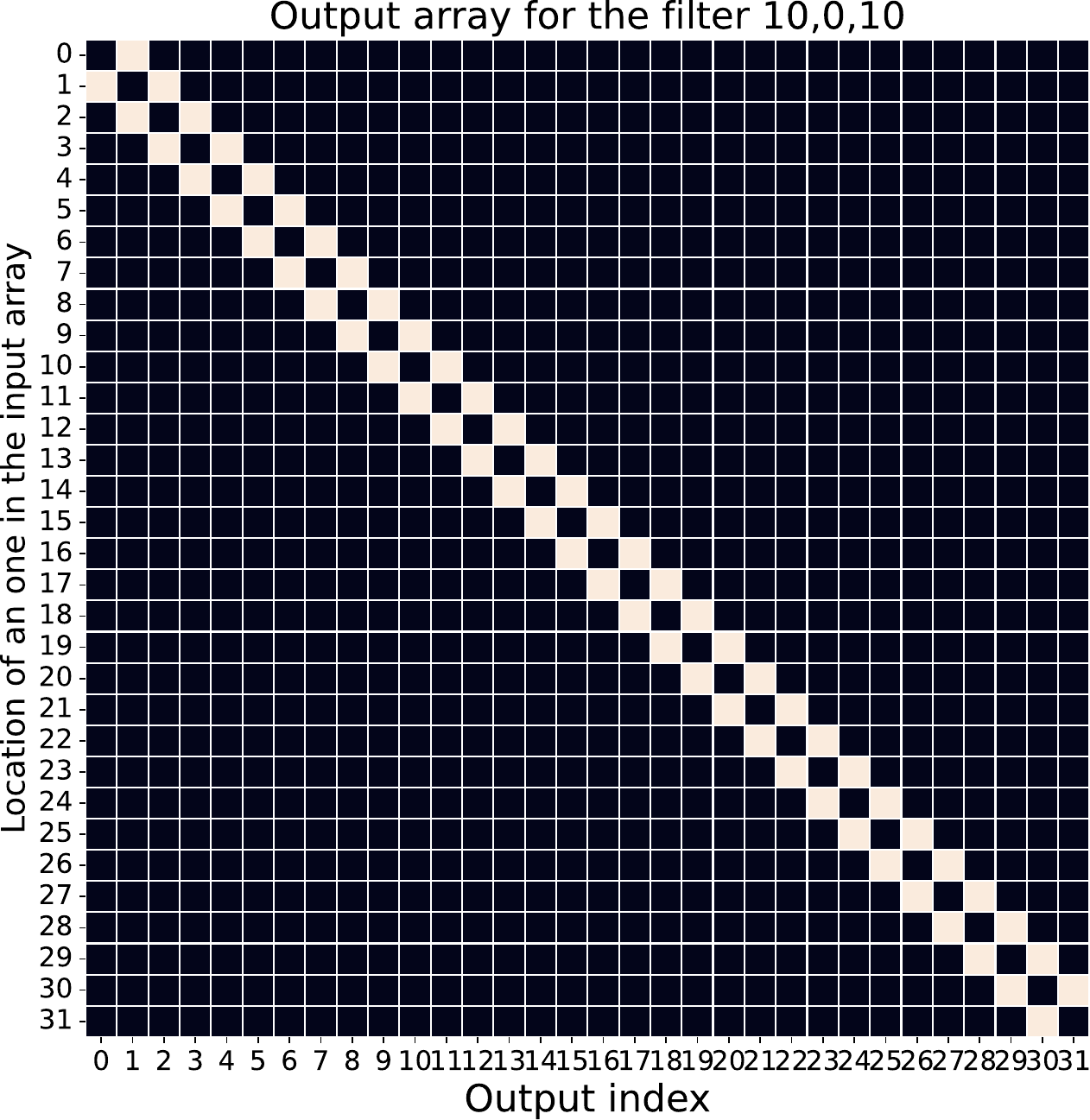}&\includegraphics[scale=0.1]{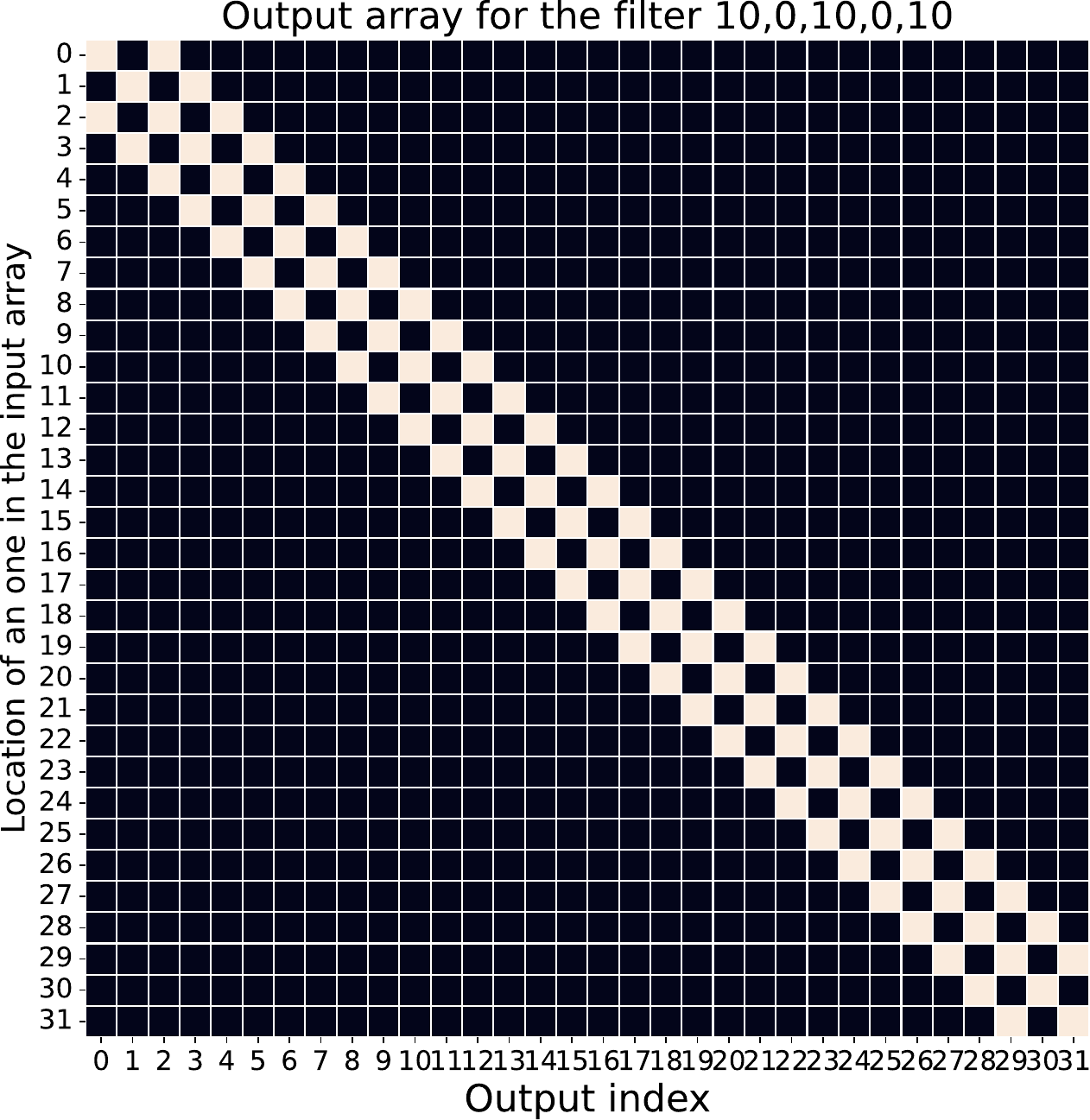}&\includegraphics[scale=0.1]{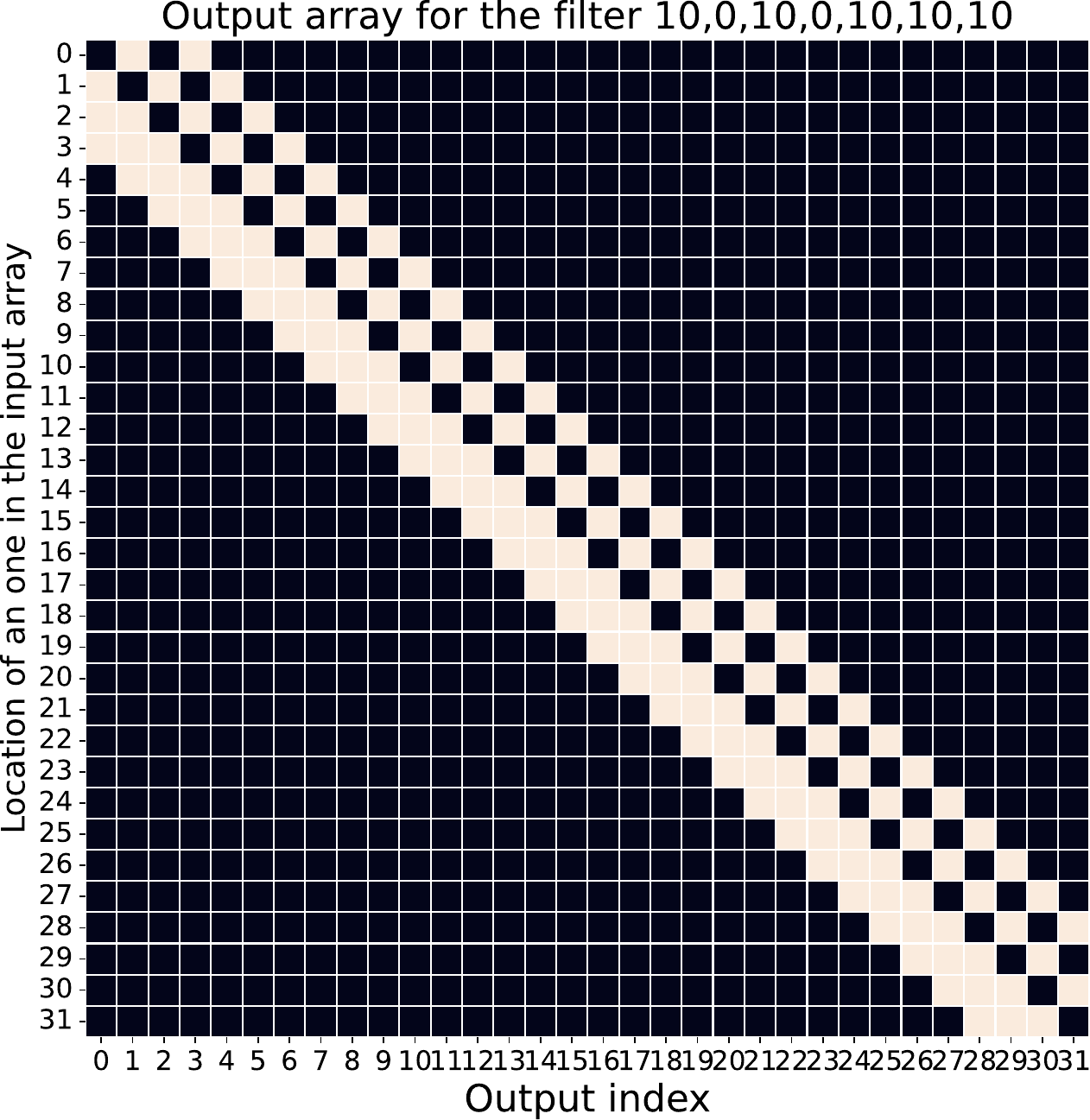}&\includegraphics[scale=0.1]{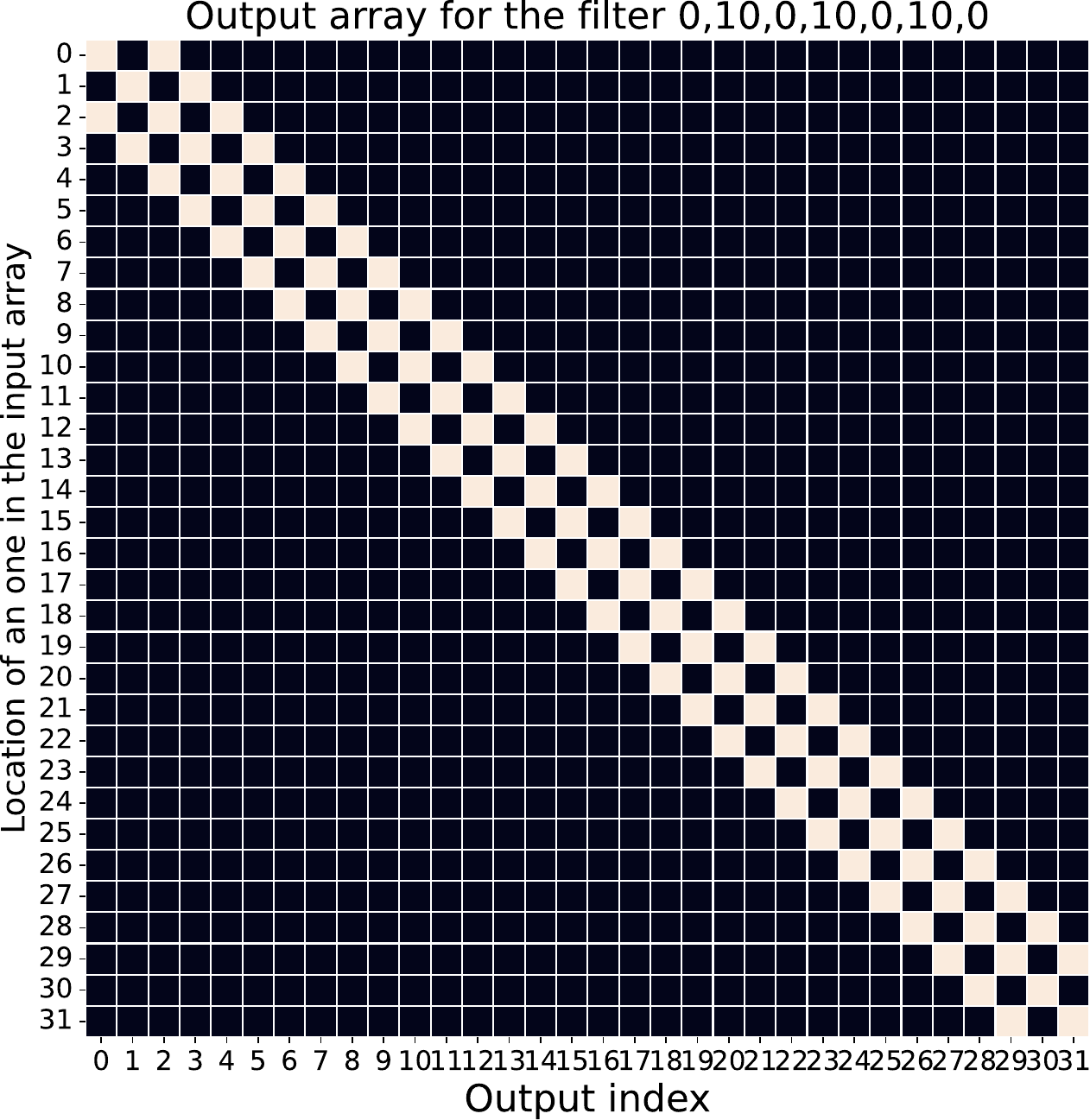}
\\

\end{tabular}

\figcaption{Illustration of the {\em HuffDuff} attack on 1-D convolution. The first row shows the number of elements and
sparsity present in the filter. The second row shows a heatmap of 32 outputs (with 32 elements each) obtained by
convolving different 1-D inputs with the sparse filter.}
\label{fig:patterns1D}
\end{table*}

\begin{figure*}[h]
    \centering
    \begin{tabular}{ccc}
     \includegraphics[scale=0.51]{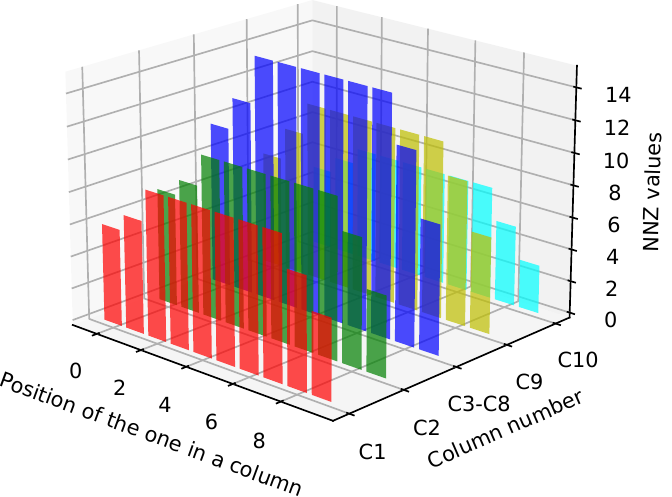} &  \includegraphics[scale=0.51]{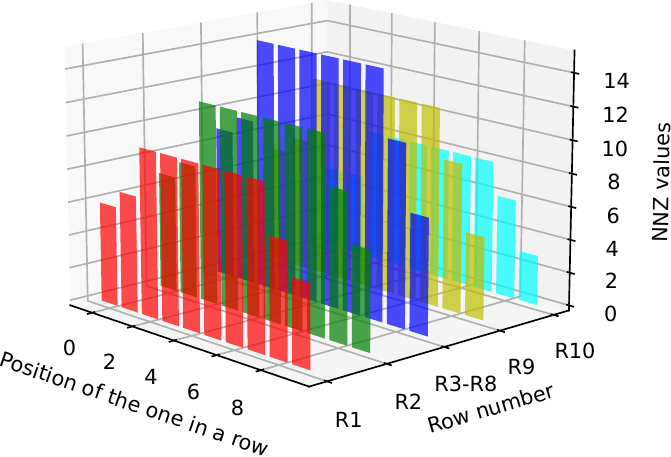} &  \includegraphics[scale=0.51]{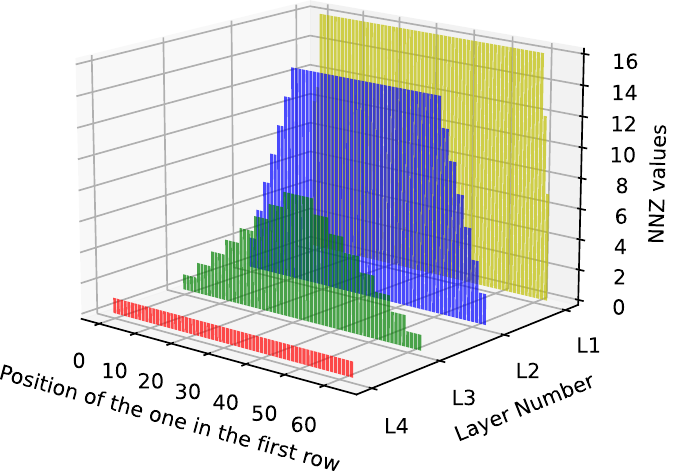}\\
     \textbf{(a)} & \textbf{(b)} & \textbf{(c)}\\
    \end{tabular}
    \caption{Illustration of the {\em HuffDuff} attack on a 2D convolution. \textbf{(a)} The number of non-zeros values
using a column-major input. \textbf{(b)} The number of non-zeros values using the row-major input. \textbf{(c)} Impact of the
boundary effect across different convolution layers }
    \label{fig:patterns2D}
\end{figure*}

%% file: threatModel.tex
\section{System Design and Threat Model}
\label{sec:threat}
\begin{figure}[!htb]
    \centering
    \includegraphics[scale=0.3]{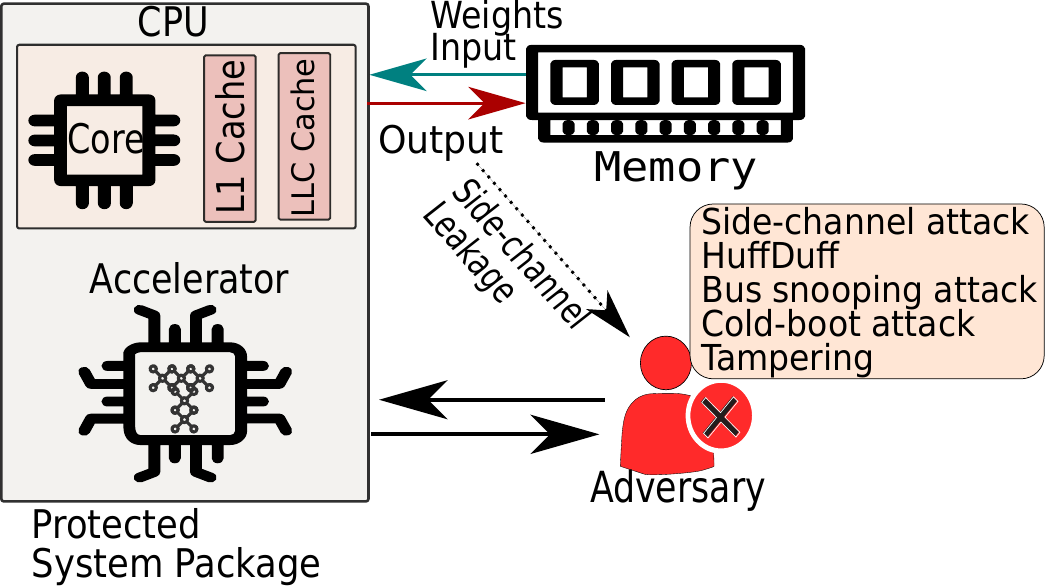}
    \caption{The system design and the threat model (similar to \cite{reverse,huffduff})}
    \label{fig:threat_model}
\end{figure}

We present the high-level system design and threat model in Figure \ref{fig:threat_model}. The design and the threat
model are similar to previous attacks and countermeasures~\cite{securator,tnpu,reverse}.

We assume that the central processing unit (CPU), accelerator, and caches are co-located within a single package,
forming a system-on-chip. The CPU can send commands to the accelerator through a secure channel. However, the
transfer of data takes place through the main memory via an unsecure channel. The CPU executes a scheduler, which is
implemented on a trusted execution environment (TEE). It synchronizes the operations of the accelerator. After receiving
a few initialization instructions from the CPU, the accelerator runs on its own. This is a standard design decision~\cite{guardnn,tnpu,securator}.
It is worth noting that the CPU, accelerator and caches are a part of 
the Trusted Computing Base (TCB). The TCB is a collection of hardware components that are considered to be secure. 

An attacker can target the system memory and memory interface. In the given scenario, potential adversaries might
include the operating system, the hypervisor, a malevolent application, or somebody having physical access to the main
memory or memory bus. The adversary has the capability to extract the memory address sequence and detect the type of
memory access instruction (whether it is a read or write operation) from the memory bus using any side channels. She can
utilize this information to determine both the model parameters and the model architecture. She can also access the
traffic and the content of the buses. For sparse NNs specifically, she can craft specialized inputs and feed them to the
NN. Note that all traffic is encrypted and data stored in the main memory (outside the TCB) is tamper proof in our
design. 

%% file: attack.tex
\section{Analysis of Attacks}
\label{sec:huffduff}
We shall only discuss Type A and B attacks in this section. Type C attacks are far simpler and there is a fair amount of work already in that area~\cite{securator,tnpu,guardnn}. 

\subsection{Analysis of {\em Type A} Attacks}
\label{sec:attack}
In this section, we discuss the basic methodology for stealing the architecture of a dense NN. An adversary probes the memory traffic via a side channel and reverse engineers the model parameters based on the leaked memory signatures. There are certain memory signatures that serve as critical hints to extract the true architectural model. The architecture of a convolution is categorized by its type, size, and the boundary between the layers. These three main identifiers are categorized using six architectural leakage hints as shown in Table ~\ref{tab:hints}. 

\begin{figure}[h]
  \centering
  \includegraphics[scale=0.14]{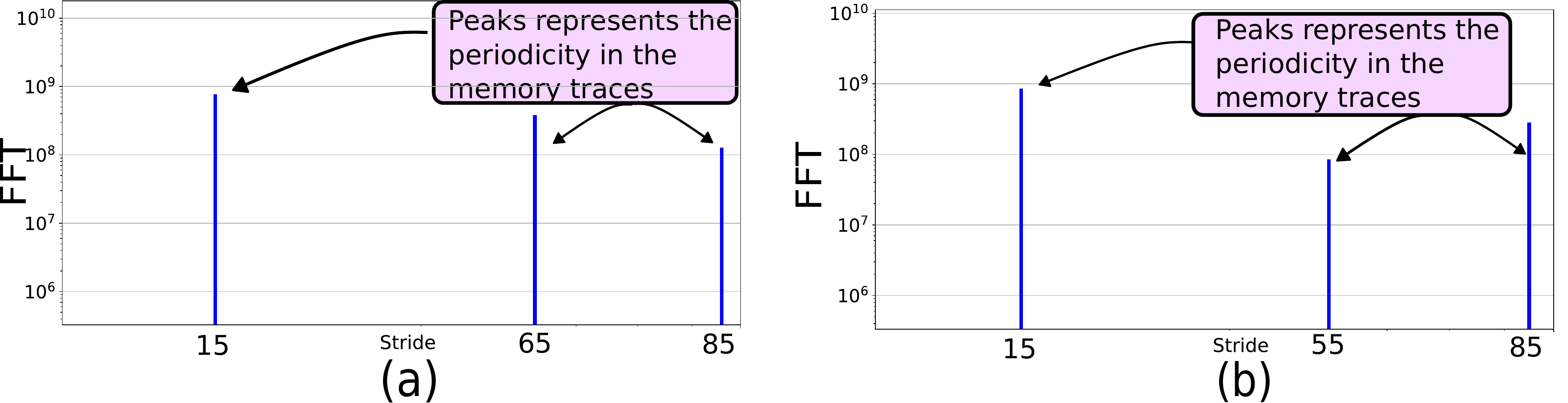}
  \caption{Attacking a dense NN by measuring the periodicity of memory accesses (using FFTs). (a) Loop stride (15,65,85) (b) Loop stride (15,55,85) }
  \label{fig:attack}
\end{figure}

In order to characterize the periodicities present in the leaked signatures, we rely on the classical {\em Fast-Fourier transform}(FFT) (a small yet novel contribution in this paper). 
We consider the signal to to be the sequence of tiles, and plot the FFTs in Figure~\ref{fig:attack} (also refer to \cite{securator} that has a similar analysis in a different form). \textbf{The FFT helps us  detect the loop sizes (layer dimensions) very easily}.  In Figure \ref{fig:attack}, we show the FFTs for the {\em Vgg16} net (Layer 1). It is clearly visible that the inputs and outputs are tiled using a three-level loop. The size of each loop corresponds to the tiling factor; this manifests as three peaks (100\% correlated with the corresponding model parameters).

\begin{table}[h]
    \centering
    \footnotesize
    \begin{tabular}{|p{0.4\textwidth}l|}
    \hline
    \rowcolor{blue!10}
    \multicolumn{2}{|c|}{\textbf{Layer boundary identifier}} \\
    \hline
     Read-after-write distance & \\
     Periodic filter access sequence &   \\
    \hline
    \rowcolor{blue!10}
    \multicolumn{2}{|c|}{\textbf{Layer dimension identifier}} \\
    \hline
    Memory accesses in each layer (periodicity) &  \\
    Identify weights and \ofmaps (former is read-only data, latter has a read-write access pattern) &   \\
    \hline
    \rowcolor{blue!10}
    \multicolumn{2}{|c|}{\textbf{Layer type identifier}} \\
    \hline
     Memory access patterns in various layers (periodicity) &  \\
    \hline
    \end{tabular}
    \caption{Architectural hints for estimating the architecture of a dense model}
    \label{tab:hints}
\end{table}

%% file: characterization_attack.tex
\subsection{Analysis of {\em Type B} Attacks }

%There is a sufficient amount of scholarship on Type A and Type C attacks~
%\cite{}; hence, we focus on Type B attacks, where
%characterization studies are eponymously sparse. 

Dignqing et al.~\cite{huffduff} proposed a novel attack {\em HuffDuff} for sparse accelerators. The key insight is that due to the boundary effect in convolution, different \ofmaps have different compression ratios (read non-zero(NNZ) elements).
By providing a sequence of well crafted inputs (essentially 2D impulse functions), an adversary observes the impact on the number of non-zero (NNZ)
values. Using this information, it is possible to figure out the filter sizes (primarily) across different layers of a large network.  

\subsection{Attacking a 1D Convolution Layer} For an input size of
$32$, we require 32 different \ifmaps, which place the `1' value (rest 0s) at different positions.  For different
filter sizes [3,5,7] and different levels of filter sparsity, different \textbf{patterns} are observed (see
Figure~\ref{fig:patterns1D}). We observe that the pattern of the NNZ values in the \ofmaps changes with the filter
dimension and sparsity -- there is a discernible correlation, which can be used to reliably guess filter sizes.

\subsection{Attacking a 2D Convolution Layer} Consider an image with the dimensions $10\times10$. Here also, we move one
non-zero (NZ) pixel across the image, while all the other image pixels are set to zero. Figure (a) in Figure~\ref{fig:patterns2D}
shows the \textbf{patterns of NNZ values} in the \ofmaps as the NZ value is moved across the different positions in
ten different columns of the input image. Figure (b) shows the result of repositioning the NZ value across different
positions in the 10 different rows of the \ifmap. We observe that the boundary effect is more pronounced at the edge than at
the center-most row/column.  Next, we analyze the impact of an attack on various CNN layers using a scaled version of LeNet\cite{lenet} (5 layers).
The image (c) in Figure~\ref{fig:patterns2D} illustrates that the boundary effect is most pronounced at
the first layer and decreases as we move deeper into the layer. Nevertheless, we can see that some information is {\em being leaked}. We can use Equation~\ref{equ:boundary} to verify that the filter sizes can indeed be computed from this data -- the most important feature is the point at which each graph becomes a straight line (e.g. 3 in image (a) for C1), 7 in image (c) for L3).

%\subsection{Summary} 

%\noindent $\blacktriangleright$ {\em Patterns of NNZ values}: The patterns of these NNZ values
%allow an adversary to reverse-engineer the dimensions of the filter. We shall later
%conduct {\em Runs}~\cite{runstest} and {\em
%Cramer-von Mises}~\cite{CVM3} tests to show that \fname converts these patterns into a random distribution (see
%Section~\ref{}).  

%\noindent $\blacktriangleright$ {\em Boundary effect across layers}: It is more pronounced in the initial layers %of a
%CNN as the receptive fields of neurons near the border are smaller and contain less spatial context. It tends to %diminish as we progress deeper into
%the network due to the use of techniques such as padding and skip connections, which mitigate the loss
%of spatial information. 

%Consequently, they
%may be affected by a reduced number of input values, resulting in %more pronounced boundary effects.

%\noindent $\blacktriangleright$ {\em Deeper layers}: The boundary effect tends to diminish as we progress deeper into
%the network. This is the result of the use of techniques such as padding and skip connections, which mitigate the loss
%of spatial information and enhance network performance. 
%These techniques improve the propagation of information between
%the input and output layers, thereby minimizing the boundary effect.

%The intuition is that the boundary effect can survive multiple layers, but as it travels through more layers, the
%footprint of the probe impulse (the ``active'' position of the `1' in the crafted input) gets increasingly diffused.

\subsection{Limitations of State-of-the-art CMs} 

The seminal work by Li et al.~\cite{neurobfuscator} proposed a set of
obfuscation strategies to conceal memory access patterns. They propose to widen the convolution layer by zero padding,
increase the dimension of the kernel, padding the input, or fusing/splitting layers. All of these methods increase the
computation overhead substantially, and we shall see do not comprehensively hide all access patterns such as read-after-write
dependences (also pointed out in~\cite{dnncloak}). The number of NNZ values also don't change, and thus later layers that
are often sparser~\cite{ebpc} are prone to HuffDuff style attacks.

Some authors propose encrypting the crucial weights
of an NN layer~\cite{sealing,dnncloak} to hide side-channel information. The NNZ values in the \ofmap will be
\textbf{unaffected by the encryption of the weights}. Consequently, this line of CMs is also not useful
against {\em HuffDuff}.

{\bf Our aim is to show using rigorous statistical and information theoretic measures that we effectively hide all the side-channel
based information that have been used in prior work (the most important ones were discussed in this section)}.

%% file: characterization.tex
\section{Characterization of Compression Algorithms}
\label{sec:char}

\subsection{Setup and Benchmarks}

We characterize the behavior of state-of-the-art compression algorithms using the setup mentioned in Table
\ref{tab:setup}. We used pruned NNs as the benchmarks (same as our nearest competitor \cite{huffduff}). The models are
generated using the layer-adaptive magnitude-based pruning scheme~\cite{lamp} proposed by Lee et al.~\cite{lamp}.
 The scheme achieves an optimal trade-off between performance and sparsity. We use the NVIDIA Tesla
T4 GPU with CUDA 12.1 to train and prune the NN models.  After training the models using the CIFAR-10 dataset, the
pruning algorithm is applied. We then fine-tune the pre-trained and pruned model to improve its precision. We achieved
an average accuracy of 90\% for {\em Vgg16} and 89\% for the {\em ResNet18} -- same as the original authors.

\begin{table}[!h]
    \footnotesize
    \centering
    \begin{tabular}{|p{35mm}|p{43mm}|}
    \hline
   \rowcolor{blue!10}
    \multicolumn{2}{|c|}{\textbf{Hardware Settings}}  \\
    \hline
    Intel Core i5-8300H CPU, 2.3 GHz & DRAM: 8 GB \\
    \hline
    \#CPUs: 8 Cores & L1:128 KB, L2:1 MB, L3:8 MB \\
    \hline
    \rowcolor{blue!10}
    \multicolumn{2}{|c|}{\textbf{Software Settings}} 
    \\
    \hline
    Linux kernel: 5.15 & Operating System: Ubuntu 20.04 \\
    \hline 
    Python version: 3.8 & Pytorch version: 1.11 \\
    \hline
    GCC: 9.4 & Boost version: 1.71 \\
    \hline
    \hline
    \end{tabular}
    \caption{System configuration }
    \label{tab:setup}
\end{table}

\subsection{Performance Insights}
\begin{figure}[!htb]
    \centering
    \includegraphics[scale=0.32]{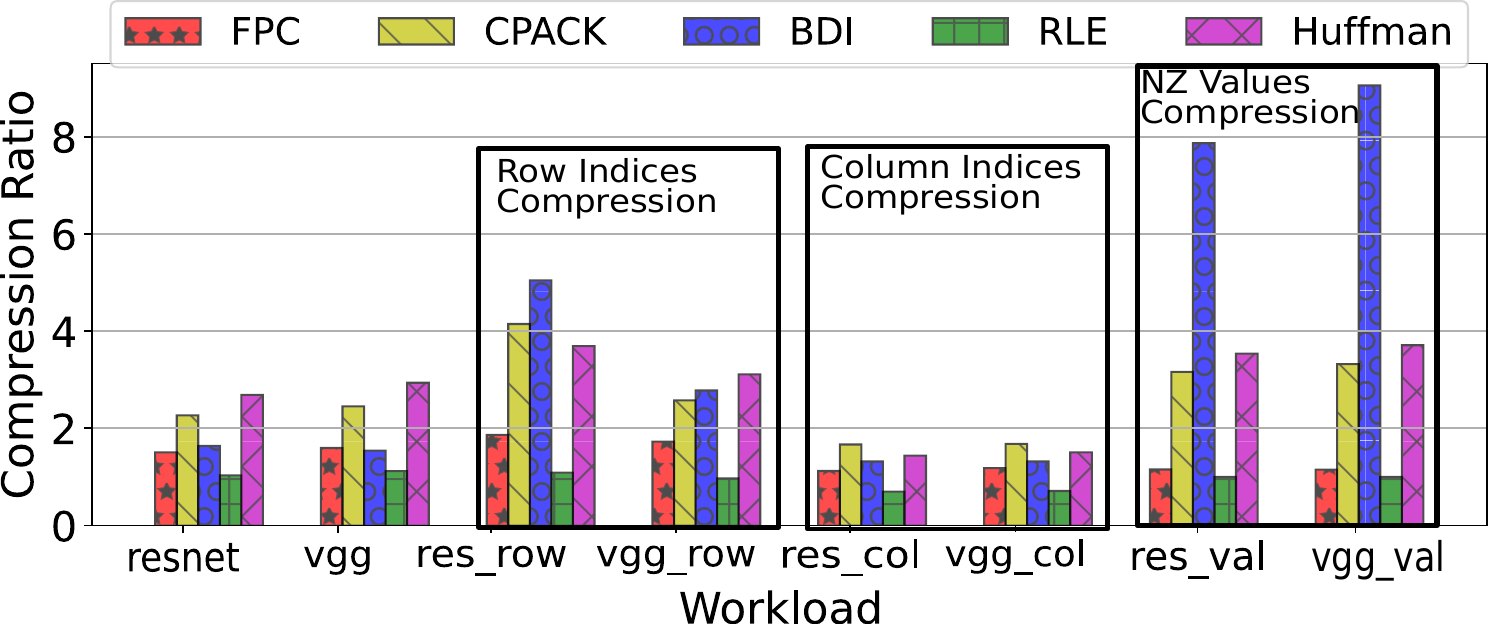}
    \caption{Characterization of the compression ratio for different compression algorithms}
    \label{fig:CR}
\end{figure}

\begin{figure}[!htb]
    \centering
    \includegraphics[scale=0.32]{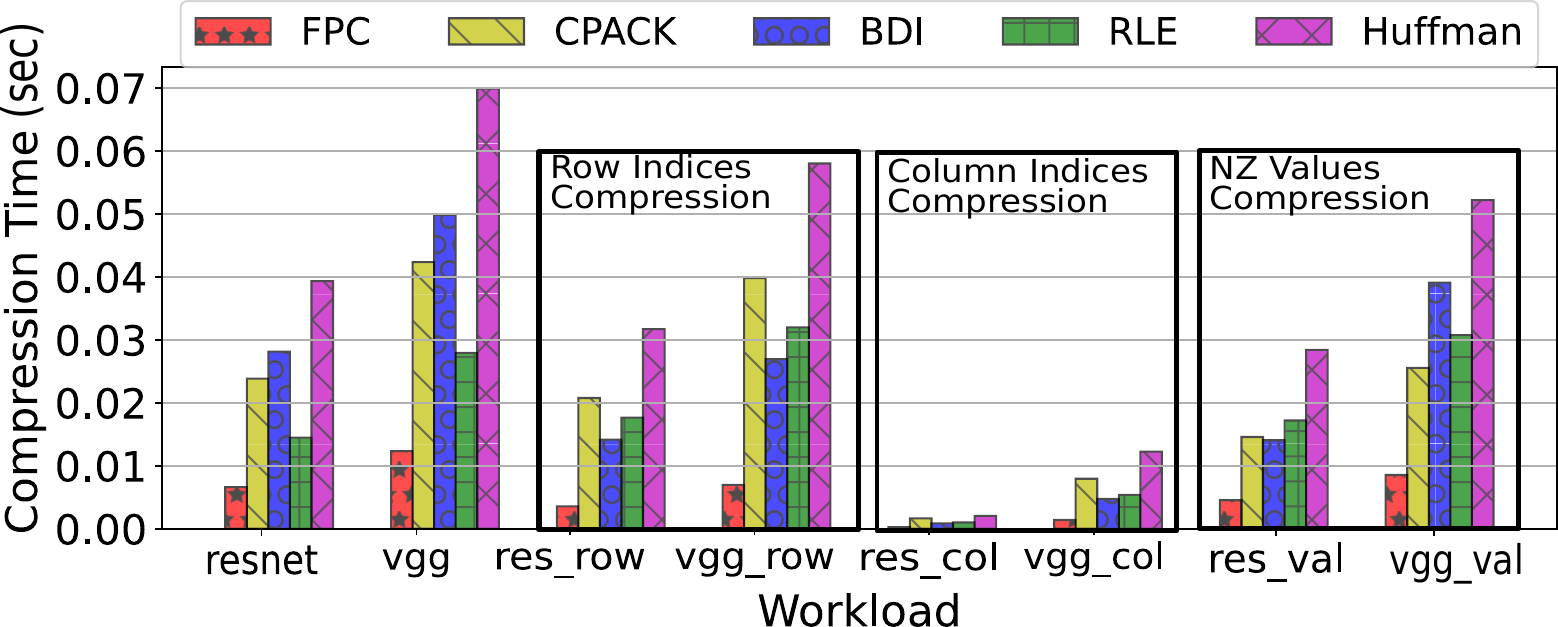}
    \caption{Characterization of the compression time for different compression algorithms}
    \label{fig:CT}
\end{figure}

We use different compression algorithms to compress the NNs. We present the compression ratio as well as the compression
time in Figures ~\ref{fig:CR} and ~\ref{fig:CT}, respectively. We observe that the {\em Huffman} compression provides
the best compression ratio; sadly, its compression time is poor as compared to the other hardware-based compression
algorithms. The BDI algorithm provides an optimal trade-off between the compression ratio and compression time. 

\begin{figure}[!htb]
    \centering
    \includegraphics[scale=0.21]{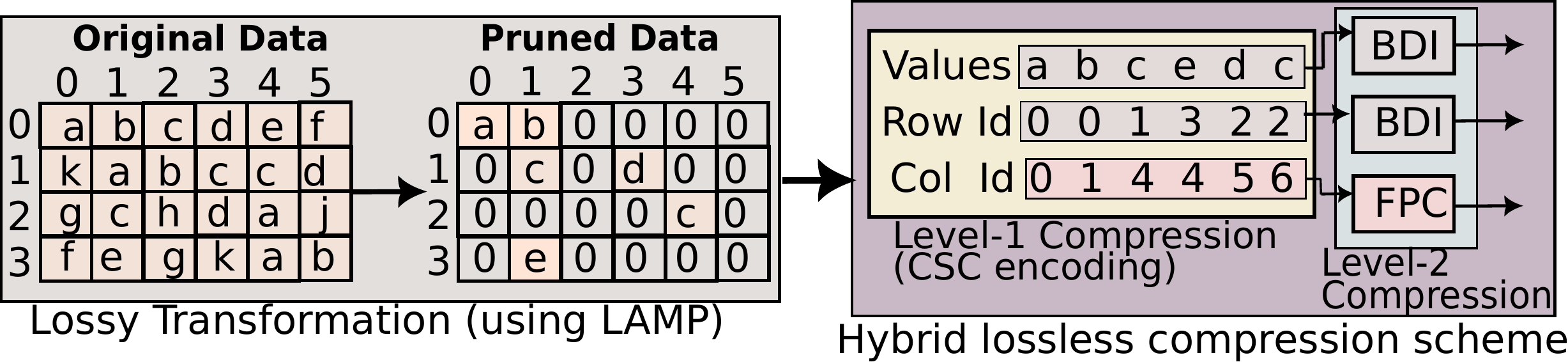}
    \caption{Compression scheme for sparse NNs. Level-1 encodes the sparse data, while Level-2 further compresses the encoded data.}
    \label{fig:CSC}
\end{figure}

In order to further improve the compression process, we encoded the models in the compressed sparse column (CSC) format, as
depicted in Figure \ref{fig:CSC} (same as Eyeriss v2\cite{eyerissv2}). This sparse encoding reveals the
patterns in the data and is extensively used to represent data in a host of sparse accelerators\cite{eyerissv2,scnn}.
The sparse data is represented by {\em arrays} of NNZ values, row IDs, and col IDs. The row ID is the row
number of the NNZ value within a specified array. The column ID indicates the number of consecutive entries in the NNZ
array that belong to a given column. 

%\fixme{Thoroughly verify the next paragraph. I have changed parts of it. I may %have introduced errors. I don't believe
%that the values in a layer are roughly similar to each other. Really ??? Cite the figures here: fig:CT and fig:CR}
%, as the weight values are typically in the same ballpark in an NN layer %Cavigelli et al.~\cite{ebpc} and Xi. et al.~\cite{cpack} present an analogous argument in their work.  These NNZ values are very similar, as the model weights in a given layer of an NN are typically very similar~\cite{ebpc}.

We perform characterization once more to discover the best compression strategy for further compressing these arrays. We
conclude that \circlenew{1} The NNZ values are compressed most effectively by the BDI algorithm. 
\circlenew{2} We also note that for column indices, FPC is the optimal solution because it is the fastest. It does not perform well with
the other data sets, but it performs well for the column indices. \circlenew{3} We note that the compression ratio for the
BDI algorithm is $2.34 \times$ higher for row indices as compared to {\em Huffman}. Based on these insights, we design a
novel compression algorithm as shown in Figure \ref{fig:CSC}.

The main advantage of this scheme is that in the Level-2 compression, since the three compression operations (compression using the two BDI units and a  FPC unit) are independent, they can be executed
concurrently and we can design a highly parallel compression engine. We observe that the proposed scheme can generate a
compression ratio $2\times$ higher than the BDI algorithm in nearly the same amount of time. 

%\fixme{How is it parallel? Isn't it sequential. First bring into a compressed format, and then use a compression algorithm.}

\textbf{Summary: We conclude that using a hybrid compression methodology that leverages the inherent patterns within the
data generated by encoding the data in a specific format is more efficient than relying solely on basic hardware-based
compression algorithms.}

%% file: design.tex
\section{Design of \name}
\label{sec:hw}

\subsection{Overview}

\begin{figure}[!htb]
    \centering
    \includegraphics[scale=0.46]{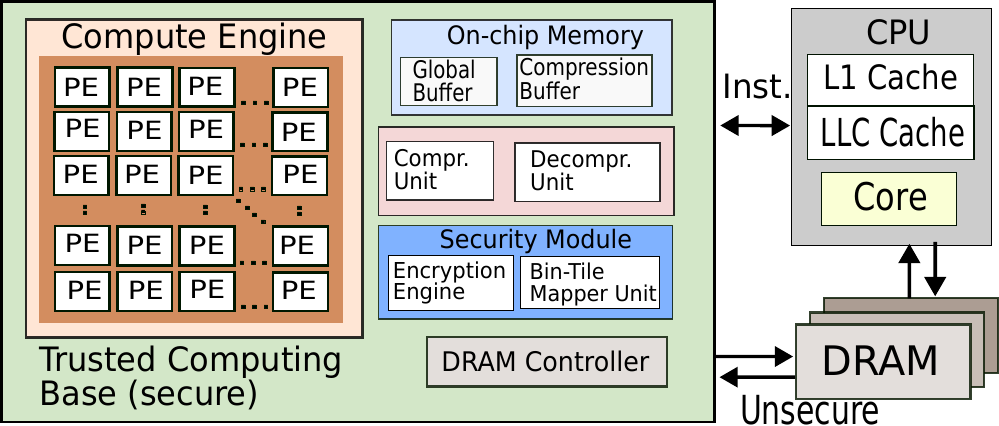}
    \caption{A high-level design of \fname}
    \label{fig:scheme}
\end{figure}

\fname provides a hardware-assisted leakage-proof secure environment for the execution of NNs. A high-level
design of the framework is presented in Figure \ref{fig:scheme}. The host CPU securely delivers instructions to the
accelerator via a PCIe link to execute a layer of the convolution. Thereafter, the accelerator starts the tile-wise
execution. All the tiled weights and inputs are compressed, encrypted and stored in their respective bins. These bins
are equal-sized contiguous blocks of memory. The location of a tile within a bin is stored in a Tile-map-Table (TMT). The
encrypted TMTs for the inputs and weights are sent to the accelerator where they are decrypted using a special session key
(provided by the user) or any other equivalent mechanism. 

During the execution of a layer, to identify which bin contains the tiles that must be processed, the scheduler 
first looks up the TMT. In the event that a tile is found within a bin, the entire contents of the bin are read and
stored within the compression buffer. If the bin is only partially full, fake read accesses  are made to make sure
that an attacker observes a full bin being read. {\em This guarantees that reads are at the granularity of a bin all
the time.} The
decompression unit again uses the TMT in order to determine the exact location of the desired tile within the bin. The
required compressed tile is decompressed by the decompression unit and stored in the global buffer before being
transmitted to the compute engine for convolution. {\bf Key point: A bin is our atomic unit of data transfer
between the secure and unsecure domains.}

The compute engine consists of a systolic PE array;   each PE has
local storage in the form of register files. Upon the completion of processing a tile, the resulting output tile
undergoes compression followed by encryption. The compressed and encrypted output data is stored in the global buffer. A
Bin-Tile mapper unit determines whether the tile can be stored in the current bin and updates the TMT to indicate which
tiles are stored in which bins.  Once a bin reaches its maximum capacity, meaning that the next compressed and encrypted
tile cannot be accommodated, all the contents of the bin are written to the main memory. Even if the bin is only
partially full, we pad the rest of the bin with zero values and encrypt them using a random key. The bin is then {\bf encrypted} and written to the main memory.
{\em This ensures that writes also are at the granularity of bins.}  Each \ofmap is buffered in the on-chip memory, similar to
~\cite{dnncloak}. {\em This not only obfuscates the data traffic but also breaks RAW dependencies in the data access
patterns.}
  
It is crucial to remember that the user determines the bin size, which is independent of the model architecture. In
the following sections, we will provide details of the compression and binning strategies.  

\input{BPP}

%% file: BPP.tex
\subsection{Compression Scheme}
We compress the \ifmaps, \ofmaps and weights using a compression engine as shown in Figure \ref{fig:comp_HW}. We cannot use CSC to represent non-sparse data because doing so would increase the data
size rather than decrease it. Thus, we bypass the CSC engine and re-use the BDI engine for dense accelerators. This
ensures that the proposed CM is applicable to both sparse and dense accelerators. The encoding process will
not add any extra cycles since the compression engine can operate simultaneously when PEs process other sets of data.
Other authors \cite{criticalpath,scnn} have also used a similar insight, albeit in a different
context. 

Each compressed tile is encrypted using two parallel, pipelined AES-CBC-256 engines. The encryption process ensures
that the NNZ values present in the \ofmap are obfuscated. Then, the bin-tile mapper unit assigns each compressed and
encrypted tile to a bin.

\begin{figure}[!htb]
    \centering
    \includegraphics[scale=0.7]{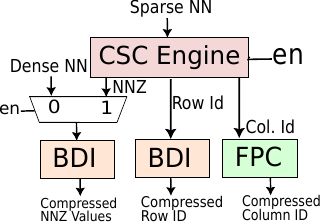}
    \caption{A high-level design of the compression engine}
    \label{fig:comp_HW}
\end{figure}

\subsection{Tile-to-Bin Mapping}

We rely on a classical bin-packing heuristic to place the tiles into the bins. \textit{Bin packing is a known method of information hiding}~\cite{sgxBPP} primarily because the inverse bin packing problem~\cite{IBPP} is NP hard; it is hard to guess what the bins contain. We shall rely on the simple next-fit algorithm to map each tile
to a bin since it does not distort the sequential nature of the data -- preserving spatial locality proves to be very beneficial from a performance perspective. We simply keep adding tiles to a bin till it is full -- a tile is never split across bins. We can of course obfuscate this process further by adding some randomization; however, this added randomization is not required in practice.

%There are certain constraints that need to be followed \circlenew{1} Ensure that the capacity of each bin is respected.
%\circlenew{2} Each compressed tile should be placed in a bin.  \circlenew{3} After each computation, the \ofmap tile is
%compressed and stored in the global buffer. This process is repeated until the bin is full or until the next compressed
%tile cannot be stored in the current bin. This generates the next constraint, which states that the bin capacity must be
%less than the size of the global buffer.
%We read and write the data from the memory at the granularity of a bin, irrespective of the fact that whether the bin is
%partially filled or completely filled. 

The mapping between bins and tiles is generated by the bin-tile mapper and this mapping is stored in a Tile-map table (TMT).
The {\em Tile ID} functions
as the designated identifier for retrieving individual entries inside the TMT, which is {\em securely saved in the TCB}. Given the highly deterministic nature of neural networks, we already know the tile access schedule and we can retrieve the TMT entry in \textbf{O(1)} time.
A TMT entry comprises two fields: the {\em bin ID} and the {\em addr} within the bin, as illustrated
in Figure \ref{fig:TMT}. Keeping track of the address of each individual tile within a bin eliminates the need to
decompress the entire bin in order to decompress a single tile.  As a result, there is a reduction in performance
overheads. 

%\begin{figure}
%   \centering
%   \includegraphics[scale=0.5]{Figures/sparsity.pdf}
%   \caption{Sparsity in a Resnet model increases with the number of layers}
%   \label{fig:sparsity}
%\end{figure}

Three distinct TMTs are utilized for storing the mappings for \ifmaps, \ofmaps, and weights. The entries for each TMT
corresponding to the \ifmaps and weights are typically destroyed once a layer has been processed. The \ofmaps TMT for the previous layer acts as
the \ifmaps TMT for the next layer. We empirically estimated that the average storage space for TMTs is between $3.5-5$ KB. The modification of TMT entries can only be performed by the bin-tile mapper and the
scheduler.

\begin{figure}[!htb]
    \centering
    \includegraphics[width=0.88\columnwidth]{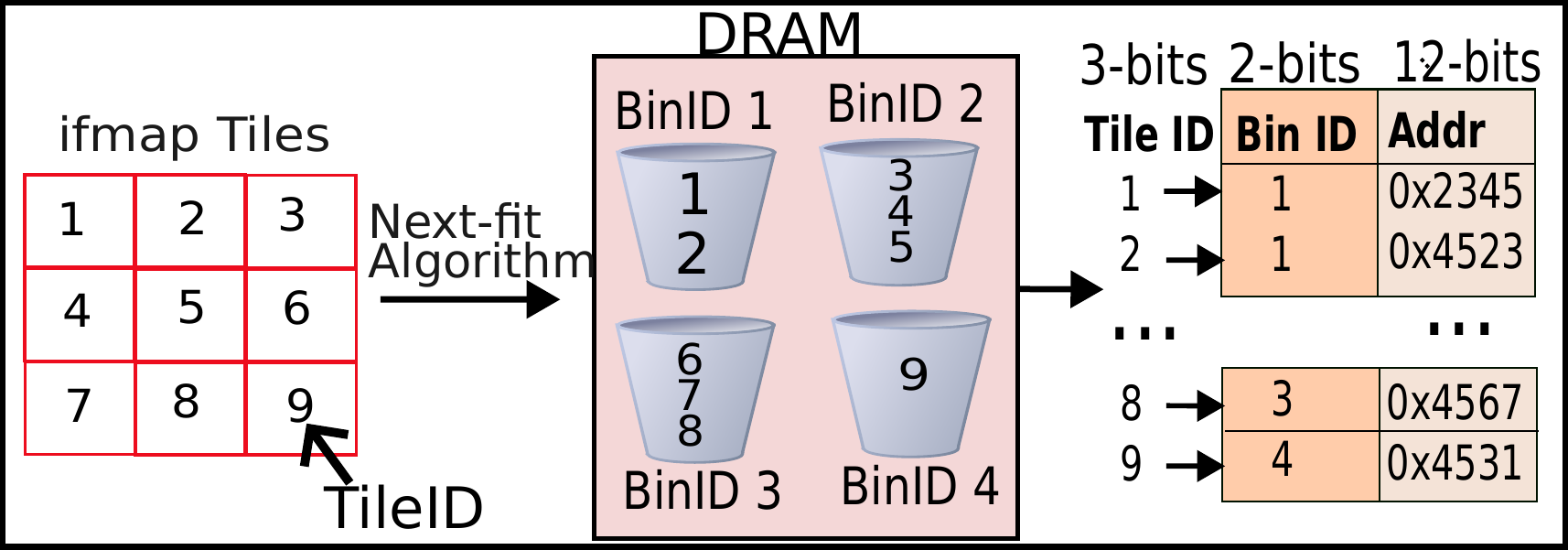}
    \caption{Tile-map table}
    \label{fig:TMT}
\end{figure}

\subsection{Data Security Unit (Securator)}
%\fixme{Rewrite this part. First para: Describe the essence of Securator. Second para: Say exactly how you will use
%Securator in our setting.}

Securator~\cite{securator} allocates a VN (version number) to every tile. The VN is incremented as the tile leaves the TCB. These VNs are generated on the fly based on the deterministic memory access patterns of a NN. It also maintains a read MAC (message authentication code) and a write MAC using these VNs to track all reads and writes within a layer. If the MACs are the same, then the data has not been tampered with. We use this mechanism as is.

\fname relies on compression and binning to obfuscate the highly deterministic memory access patterns and thus is orthogonal to Securator. It is easy to conclude that our compression and decompression units continue to process the data according to the actual tile (and VN) schedule. This sequence is indistinguishable from a regular accelerator that doesn't use \fname. So, instead of incrementing the VNs when a tile leaves the TCB as done by Securator, the VN
is incremented when a tile is compressed (write MAC is
updated). Similarly, the read MAC is updated when the tile is
decompressed. In this way, the un-obfuscated memory access
schedule can be tracked and we can reuse the Securator’s
methodology to ensure data security. 

 %Consequently, it is
%possible that a partially computed tile is waiting for the bin to
%get full so that it can leave the TCB along with the other fellow
%tiles. Now, before this could happen, the scheduler checks that
%the partially computed tile is already present in the buffer. So,
%it schedules it to get processed. Now, this would disrupt the
%DRAM memory access pattern and will make it obfuscated.
%So, instead of in-
%crementing the VNs when a tile leaves the TCB, the VN
%is incremented when a tile is compressed (write MAC is
%updated). Similarly, the read MAC is updated when the tile is
%decompressed. In this way, the un-obfuscated DRAM access
%schedule can be tracked and we can reuse the Securator’s
%methodology. The scheme will detect any malicious activity
%involving these compressed tiles as they are transferred off-
%chip

%% file: evaluation.tex
\section{Performance Evaluation}
\label{sec:eval}
\subsection{Setup}
Our experimental setup  integrates the state-of-the-art accelerator modeling tool (Timeloop~\cite{timeloop})
and a cycle-accurate DRAM simulator (Ramulator). Both the simulators have been validated against real-world hardware and
are widely used in the literature. Timeloop~\cite{timeloop} takes as input the details of the workload, the target
accelerator's configuration (see Table \ref{tab:acc_config}), and the data reuse strategy as the input and generates
performance statistics and traces. 

We developed an in-house trace generator tool that extracts the memory access patterns from the Timeloop's
T-traces~\cite{Ttrace} and then compresses and randomizes the tiles after reading the tiles in accordance with the
traces. The tool then generates a new sequence of memory accesses based on the compression and binning strategy.
Ramulator takes these new traces from our trace generator tool as its input and generates the performance statistics. 

\begin{table}[!htb]
    \centering
    \footnotesize
    \begin{tabular}{|l|l||l|l|}
    \hline
    \rowcolor{blue!10}
     \textbf{Parameter} & \textbf{Value} & \textbf{Parameter} & \textbf{Value} \\
    \hline
     PE size & 16 $\times$ 12 & Comp. buffer & 182 kB\\
     PE register & 440 B & Global buffer & 182 kB \\
      DRAM & DDR3 &  Frequency & 1.6 GHz \\
     \hline
    \hline     
    \end{tabular}
    \caption{Configuration of the accelerator (similar to \cite{eyeriss})}
    \label{tab:acc_config}
    \vskip -8mm
\end{table}

\subsection{Performance Analysis}
We consider pre-trained (dense) and pruned NN workloads (refer to  Table \ref{tab:pruneAcc}). Dingqing et
al.~\cite{huffduff} also used the same workloads; we observed that many other popular CNN models demonstrated a similar
behavior (not discussed due to a lack of space). Next, we generated pruned NNs with a fairly good prediction accuracy as shown
in Table \ref{tab:pruneAcc}. 

\begin{wraptable}[9]{r}{0.4\linewidth}
\footnotesize
    \begin{tabular}{|l|l|}
    \hline
    \rowcolor{blue!10}
    \textbf{Model} & \textbf{Acc.}\\
    \hline
    \rowcolor{gray!20}
     \multicolumn{2}{|c|}{\textbf{Pretrained}}\\
     \hline
      ResNet18~\cite{resnet_pretrain}   & 89.1\% \\
      Vgg16~\cite{vgg_pretrain} & 91.32\% \\
      \hline
     \rowcolor{gray!20}
     \multicolumn{2}{|c|}{\textbf{Pruned}}\\
     \hline
      ResNet18~\cite{resnet}  & 92.1\% \\
      Vgg16~\cite{vggnet} & 90\% \\
      \hline
    \end{tabular}
    \caption{Workloads}
    \label{tab:pruneAcc}
\end{wraptable}

  The baseline design of \fname (with compression and binning) is similar to the setup used in the {\em HuffDuff} paper. In \fname, we have used a bin size of 60 kB. The TCB can store three bins at any time. We estimate the number of tiles and tile dimensions for a given layer using Timeloop. We observed that the tile size varied with the layers and was rougly between 10 kB - 95 kB.
  
\begin{figure}[!h]
    \centering
    \includegraphics[scale=0.31]{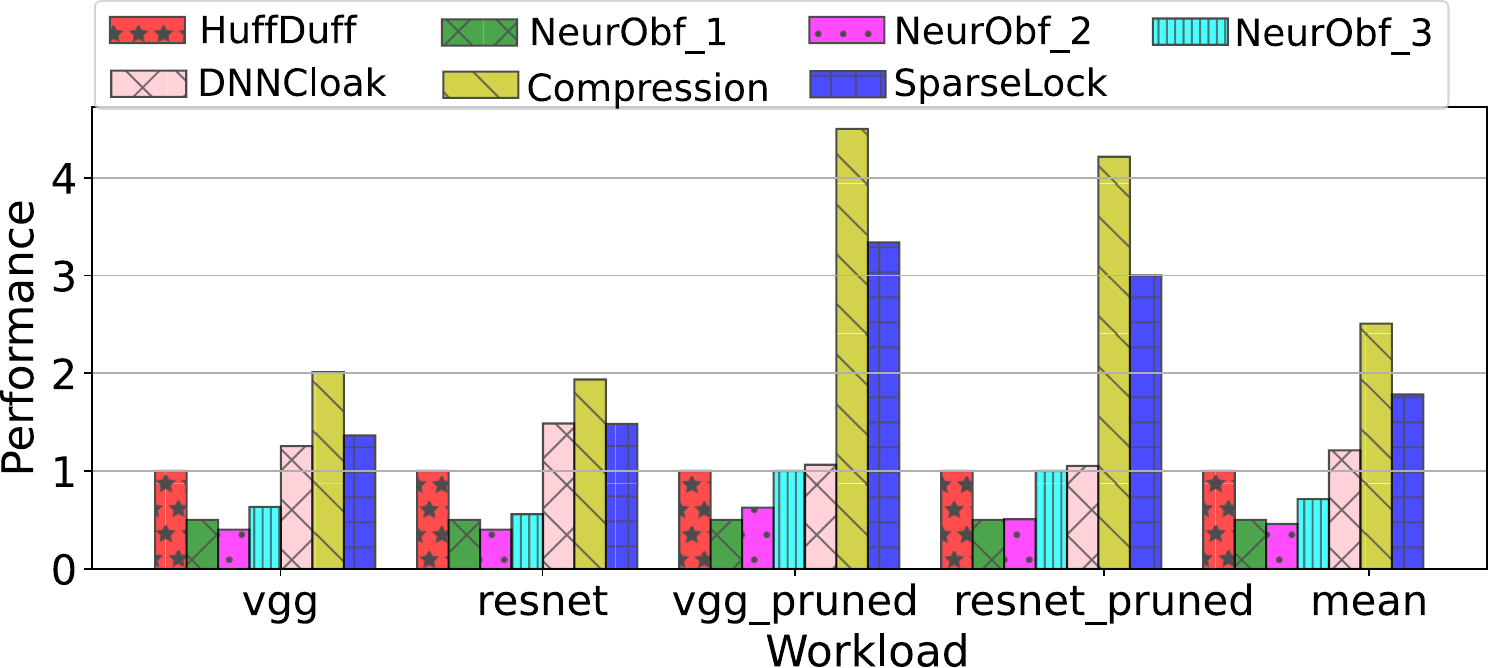}
    \caption{Normalized performance for different configurations}
    \label{fig:perf}
\end{figure}

\begin{figure}[!h]
    \centering
    \includegraphics[scale=0.31]{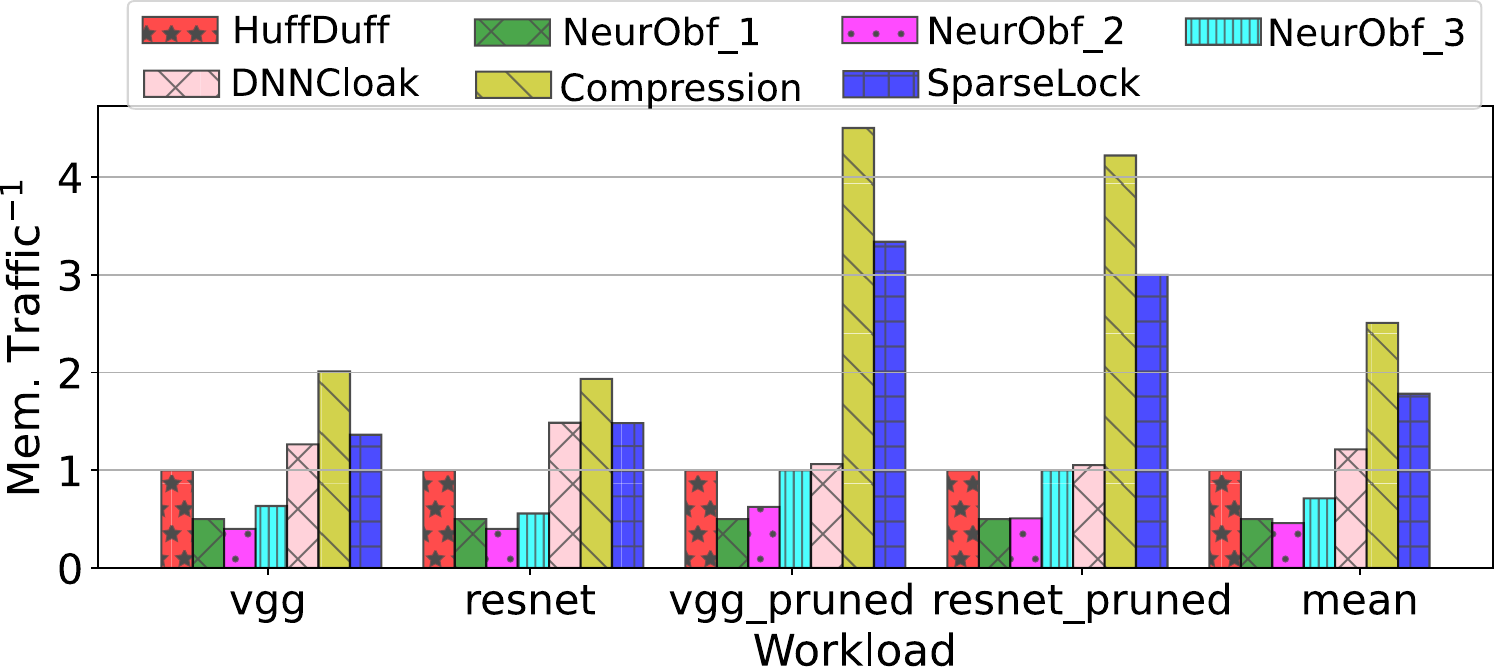}
    \caption{Normalized memory traffic for different configurations}
    \label{fig:traff}
\end{figure}

  The {\em Compr} design simply performs compressing using our proposed algorithm.  The performance results are
shown in Figure \ref{fig:perf}. The {\em performance} is proportional to the reciprocal of the total \#simulation cycles. The results are normalized to the
baseline setup (HuffDuff). We find that {\em Compr} is $2.5 \times$ faster than the baseline. This is due to
the fact that the compression of pruned NNs leads to a reduction in memory traffic by a factor of 2.5 as shown in Figure
\ref{fig:traff}. We observed a very nice correlation between performance and DRAM
traffic, which is aligned with the findings reported in prior studies ({\em TNPU\cite{tnpu}}, {\em
Securator\cite{securator}}, and {\em GuardNN\cite{guardnn}}). Next, we compared \fname, which has a 
 performance improvement equal to $1.7\times$ (vis-a-vis the baseline). 
  
We also compared our scheme with two state-of-the-art CMs -- {\em NeurObfuscator}\cite{neurobfuscator} and {\em
DNNCloak}\cite{dnncloak}. {\em NeurObf\_1} uses layer deepening, {\em
NeurObf\_2} implements layer widening, and {\em NeurObf\_3} uses kernel widening.
We see that on an average, the strategies are $2.88\times$ worse than  \fname in terms of performance. Note that
 {\em DNNCloak} performs weight
compression and incorporates an on-chip buffer to store the \ofmaps, thereby minimizing the memory traffic. But, it also
includes dummy accesses, which lead to an increase in the memory traffic, resulting in {\em DNNCloak} being $1.7 \times$
slower than \fname.

\subsection{Hardware Utilization}

The code for the compression, bin-tile mapping, and encryption units was written using the Verilog HDL. The designs were
synthesized, placed, and routed using the Cadence Genus tool for a 28 nm ASIC technology node. We present the results
in Table \ref{tab:asic}. It is evident that the overheads related to the additional hardware components are minimal.

\begin{table}[!h]
    \centering
    \footnotesize
    \begin{tabular}{|c|c|c|}
    \hline
    \rowcolor{blue!10}
    \textbf{Module} & \textbf{Area($\mu m^2 \times 10^{3}$ )} & \textbf{Power($\mu$W)}  \\
    \hline
    AES-CBC (pipelined) & 41.08  & 5.61 \\
    Compression unit & 4.82 & 0.51 \\
    Bin-time mapper unit & 0.15  & 0.20 \\
    Securator units & 1.67 & 0.22\\
   \textbf{Total} & \textbf{ 47.72} &\textbf{6.36 } \\
   \hline
   \rowcolor{gray!20}
    \textbf{Tool} & \multicolumn{2}{c|}{Cadence RTL compiler, 28 nm} \\
    \hline
    \end{tabular}
    \caption{ASIC area and power utilization for the additional hardware components used in \fname}
    \label{tab:asic}
\vskip -6mm
\end{table}

\subsection{FPGA Prototype}
We show a proof-of-concept (PoC) using the CHaiDNN framework for FPGA-based accelerators in order to estimate the
overheads (DNN compiler support and host CPU's involvement). 
CHaiDNN v2 was synthesized using Vivado HLS 2018.2 for the Xilinx Zynq Ultrascale+ ZCU102 board as the target FPGA. 
 The design required minimal modifications to incorporate all the security components and the code of the host CPU did not have to be altered.
On the basis of a dataflow generated by Timeloop, the encrypted TMT for the initial input and weights for each layer are generated. 
This data is stored as a dummy layer alongside the existing model parameters (the first layer of the DNN) by the frontend.
The design does not execute this layer and instead extracts all the information from it and populates its internal tables.

%% file: securityEvaluation.tex
\begin{table*}[!htb]
    \centering
    \footnotesize
    \begin{tabular}{c c c}
    \includegraphics[scale=0.26]{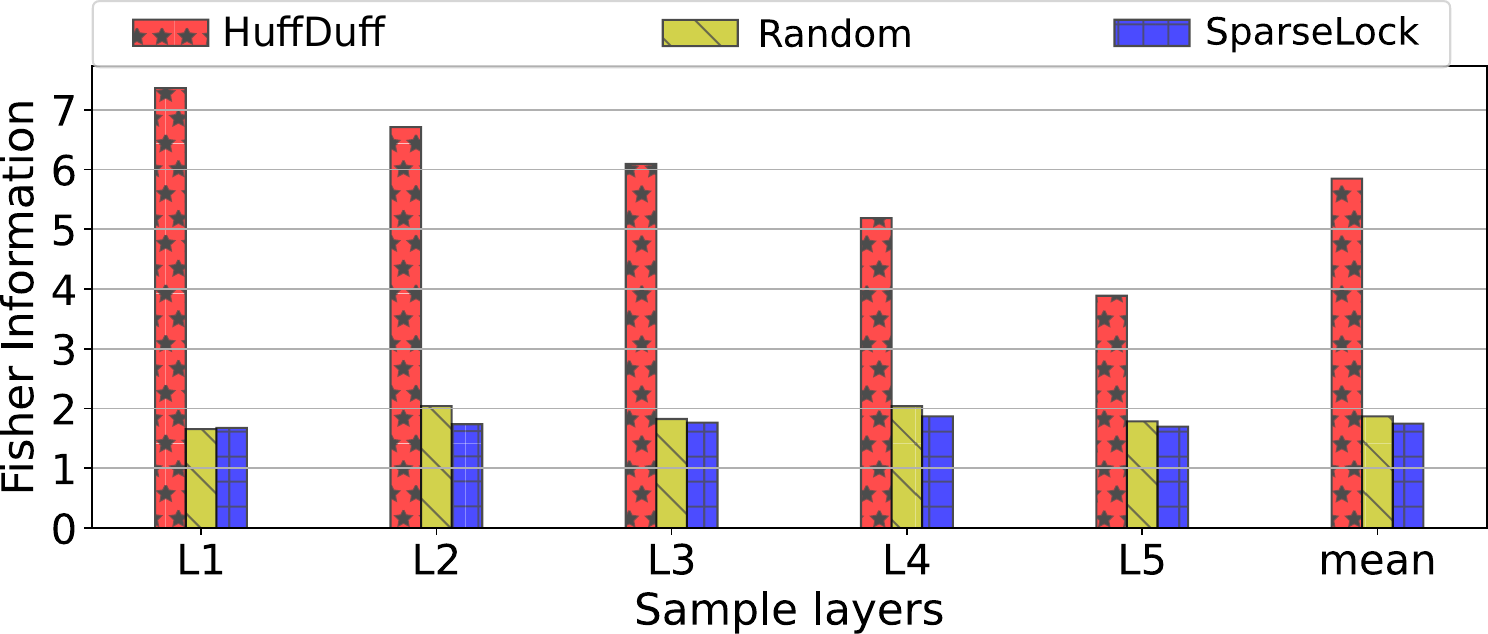}     & &\includegraphics[scale=0.26]{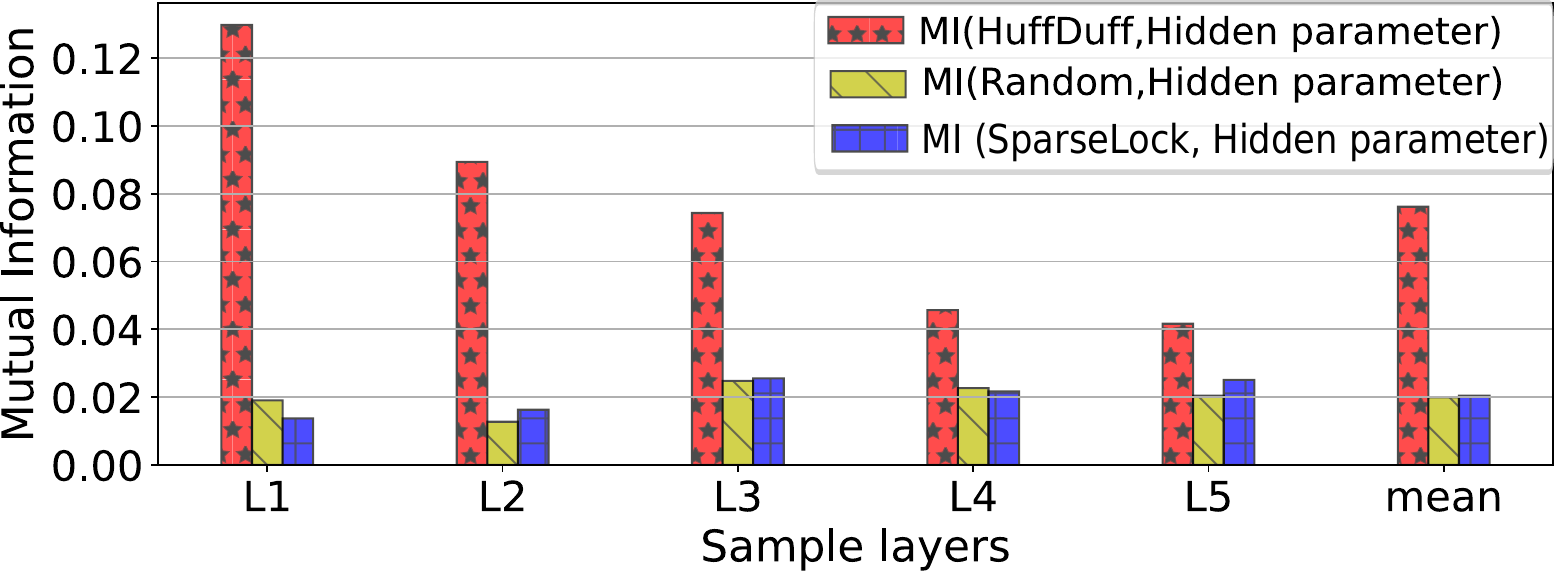}  \\
    \textbf{(a)} & & \textbf{(b)} \\
    \vspace{2.5mm}
    \includegraphics[scale=0.26]{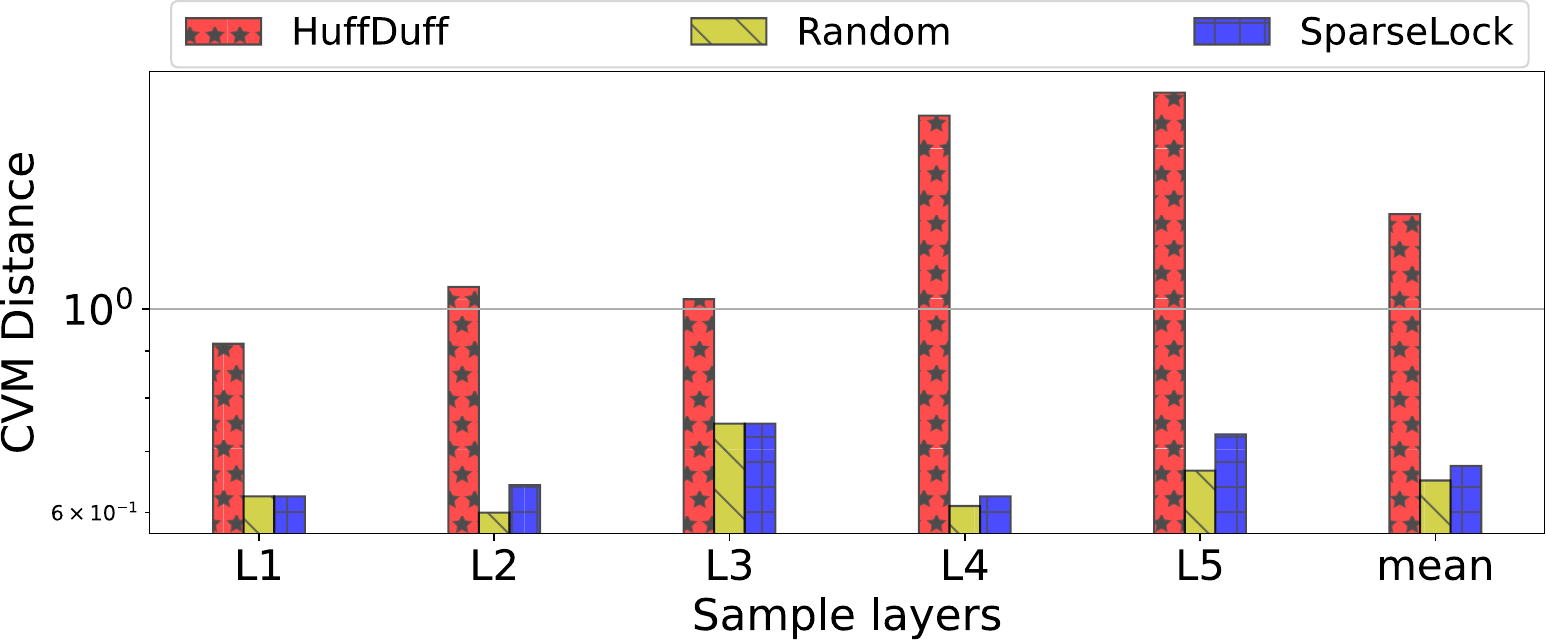}     & &\includegraphics[scale=0.26]{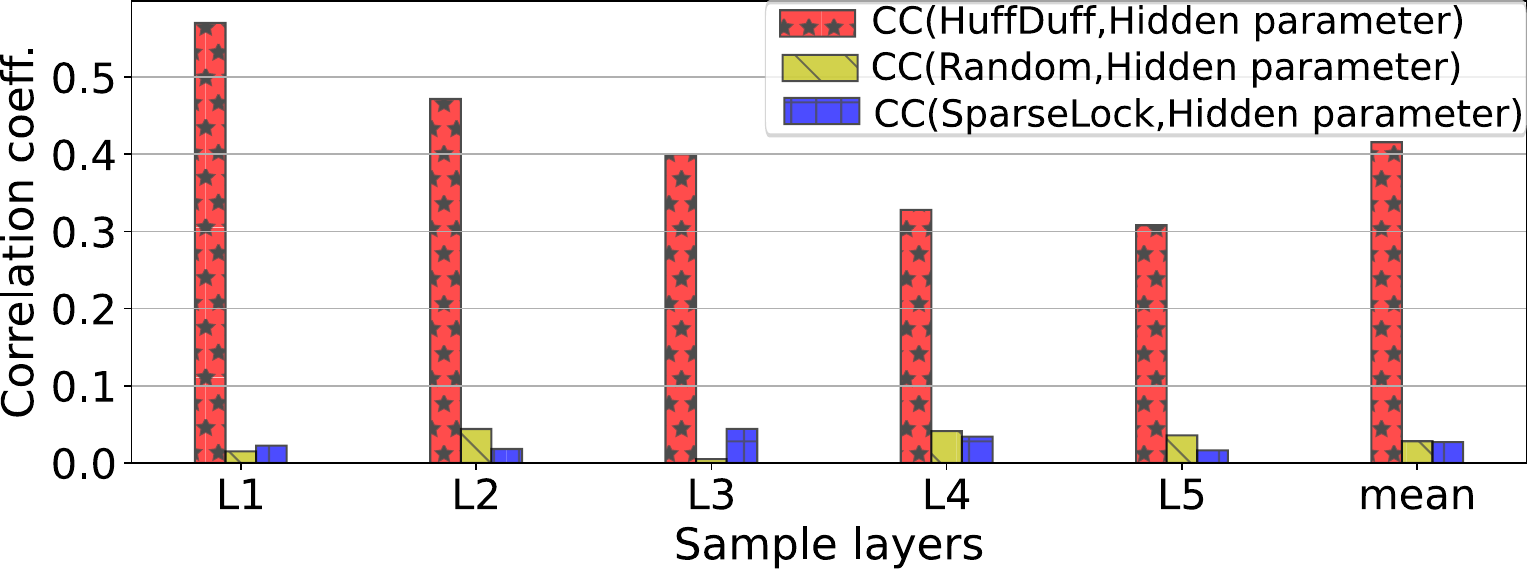} \\
    \textbf{(c)} &   & \textbf{(d)} \\
    \end{tabular}
    \figcaption{Security Analysis. \textbf{(a)} Fisher information in the leaked traces. \textbf{(b)} Mutual information
between the leaked traces and the filter size. (c) Cramer-von Mises (CVM) distance \textbf{(d)} 
Correlation coefficient between the leaked traces and the
filter size. }
\label{fig:security}
\end{table*}

\vskip -2mm 

\section{Security Evaluation}
\label{sec:security}
We utilize both information-theory-based and statistics-based tests to estimate the security of \fname. 
The cryptography community does not rely on just a single metric. For instance, the classical correlation
coefficient may be zero between the random variables, $x$ and $x^2$, even though they are clearly related.

\subsection{Defense against \HuffDuff (Sparse Accelerators)} Here, we wish to protect the NNZ values of layers 
and ensure that they don't leak out. We perform security analysis for the initial layers of the {\em
Vgg16} model for three different configurations - during the attack without any CMs (referred to as \HuffDuff), using
our proposed scheme (\fname), and using random data ({\em Random}) as shown in Figure~\ref{fig:security}. A total of 2048
distinct traces were gathered using different inputs (similar to \HuffDuff). The conclusions for other nets like Resnet18
are similar (not shown due to a lack of space).

\noindent \textbf{Information Theoretic Tests:}\\ FI (Fisher Information) and MI (Mutual Information)
have been extensively used to estimate the information
content of leaked signatures and their association with secret
data/keys~\cite{fisher1,fisher2,fisher3,fisher4,fisher5,mayer2006fisher,MIleakage,MIleakage2,MI3,MI4}. 

\noindent $\blacktriangleright$ \underline{Fisher information (FI)}  
The FI is a robust and classical method for establishing a lower
bound on the error (variance) of this estimator -- the lower the better~\cite{fisher4}. 
We mount the {\em HuffDuff} attack on various layers of {\em Vgg16} and observe that the FI
decreases as we move deeper into the layers. This is due to the fact that the quantity of information retrieved due to
the boundary effect diminishes with layer depth as described in Section \ref{sec:huffduff}. The FI in Figure~\ref{fig:security} (a) validates
the same fact.  A
similar behavior was observed in all the other layers and NNs (not presented due to space constraints). Note how close
our FI is to a truly random source (within 6.8\%).
%FIXME: XX
 
\noindent $\blacktriangleright$ \underline{Mutual information (MI)} We estimate the MI between the leaked traces and the
hidden parameter (filter size) for all the three configurations (see Figure~\ref{fig:security}
(b)). We observe a similar pattern here, indicating that the amount of {\em information between the hidden parameter and
the leaked traces for \fname is nearly the same as the information between a random distribution and the hidden
parameter.}

\noindent \textbf{Statistical Tests -}\\ For discrete samples, the runs test~\cite{runstest} (part of the NIST suite~\cite{nist}) and
CVM tests~\cite{CVM3} are considered to be the standard approaches..

\noindent $\blacktriangleright$ \underline{Runs test} The
average $p$-value for \fname is $0.365$, which is more than the standard threshold value of $0.05$ (corresponding to the
null hypothesis). This establishes the random nature of the source as per this test.

\noindent $\blacktriangleright$ \underline{Cramer–von Mises (CVM) test~\cite{CVM1,CVM2,CVM3}} We estimate the ({\em CVM distance}),
which decreases as the difference between the random data and the sampled data decreases. We observe that \fname
exhibits the lowest CVM (closest to random, see Figure\ref{fig:security} (c)).

\noindent $\blacktriangleright$ \underline{Correlation coefficient} Additionally, we also calculate the classical correlation
coefficient to determine the degree of correlation between the leaked signal and the hidden parameter using {\em
Pearson's correlation coefficient} (see Figure\ref{fig:security} (d)). Here also we have similar inferences.

\noindent $\blacktriangleright$ \underline{Architecture search space} {\em HuffDuff} significantly decreases the search
space for the $Vgg16$ and $ResNet18$ models to a narrow range of values, typically ranging from 40 to 70. \fname
expands this search space back to something akin to the original --  
$5.2\times10^{98}$ for the {\em Vgg16} model and $5.1\times10^{96}$ -- for the {\em ResNet18} model. Due to the size of the
search space, \textbf{brute force} attacks are computationally infeasible.

We would like to report that we sincerely attempted to mount the {\em HuffDuff} attack on a system that uses \fname, 
but were \textbf{unsuccessful} as the side-channel
information leaked due to the boundary effect is completely obfuscated.

\vskip -4mm
\subsection{Defence for Dense Accelerators} 
\vskip -3mm
We implement ReverseEngg\cite{reverse}, which is a state-of-the-art
memory-based SCA (side-channel attack) that steals the model architecture in the absence of any CMs; it  is the
basis for many other attacks that use the same principle such as~\cite{cachetelepathy,neurounlock,deepsniffer}. We
implemented two CMs:
({\em NeurObfuscator~\cite{neurobfuscator}} and {\em DNNCloak}~\cite{dnncloak}) -- the former introduces dummy
computations and the latter uses a combination of dummy accesses, partial encryption and extensive on-chip buffering.
We compute the distribution for the two most important architectural hints (exposed via side channels):
read-to-write distance and the total traffic distribution (see Table~\ref{tab:memAnalysis}). 
We observe minimal differences in the FI between \fname and {\em
DNNCloak} indicating that both the schemes are effective at sealing side channels. However, \fname is preferable due to
its superior performance (47.31 \% more) as shown in the previous section and DNNCloak does not protect against Type B
attacks and provides partial protection for Type C attacks.
%FIXME: XX

\begin{table}[!htb]
    \centering
    \footnotesize
    \begin{tabular}{|l|l|l|l|}
    \hline
    \rowcolor{blue!10}
     \textbf{Scheme} & \textbf{MI} & \textbf{CC} & \textbf{FI} \\
     \hline
        
     \multicolumn{4}{|c|}{Analysis of memory traffic }   \\
     \hline
     {\em NeurObfu.}~\cite{neurobfuscator} & 0.898  &  0.959   &  0.658 \\
     {\em DNNCloak}~\cite{dnncloak} &  0.846 & 0.923   & 0.038  \\
     \name   &  \textbf{0.749} & \textbf{0.566} & \textbf{0.031} \\
     \hline
     \multicolumn{4}{|c|}{Analysis of read-write distance}   \\  
     \hline
     {\em NeurObfu.}~\cite{neurobfuscator} & 0.859  &   0.807   &  1.9   \\
     {\em DNNCloak}~\cite{dnncloak} &  0.635 &  0.781  & 0.236  \\
     \name   &  \textbf{0.331} & \textbf{0.400 } & \textbf{0.103}  \\
     \hline
    \end{tabular}
    \caption{Security analysis for dense accelerators}
    \label{tab:memAnalysis}
    \vskip -3mm
\end{table}

\vskip -4mm
\subsection{Data Security: Type C Attacks}
\begin{figure}[!htb]
    \centering
    \includegraphics[width=0.8\columnwidth]{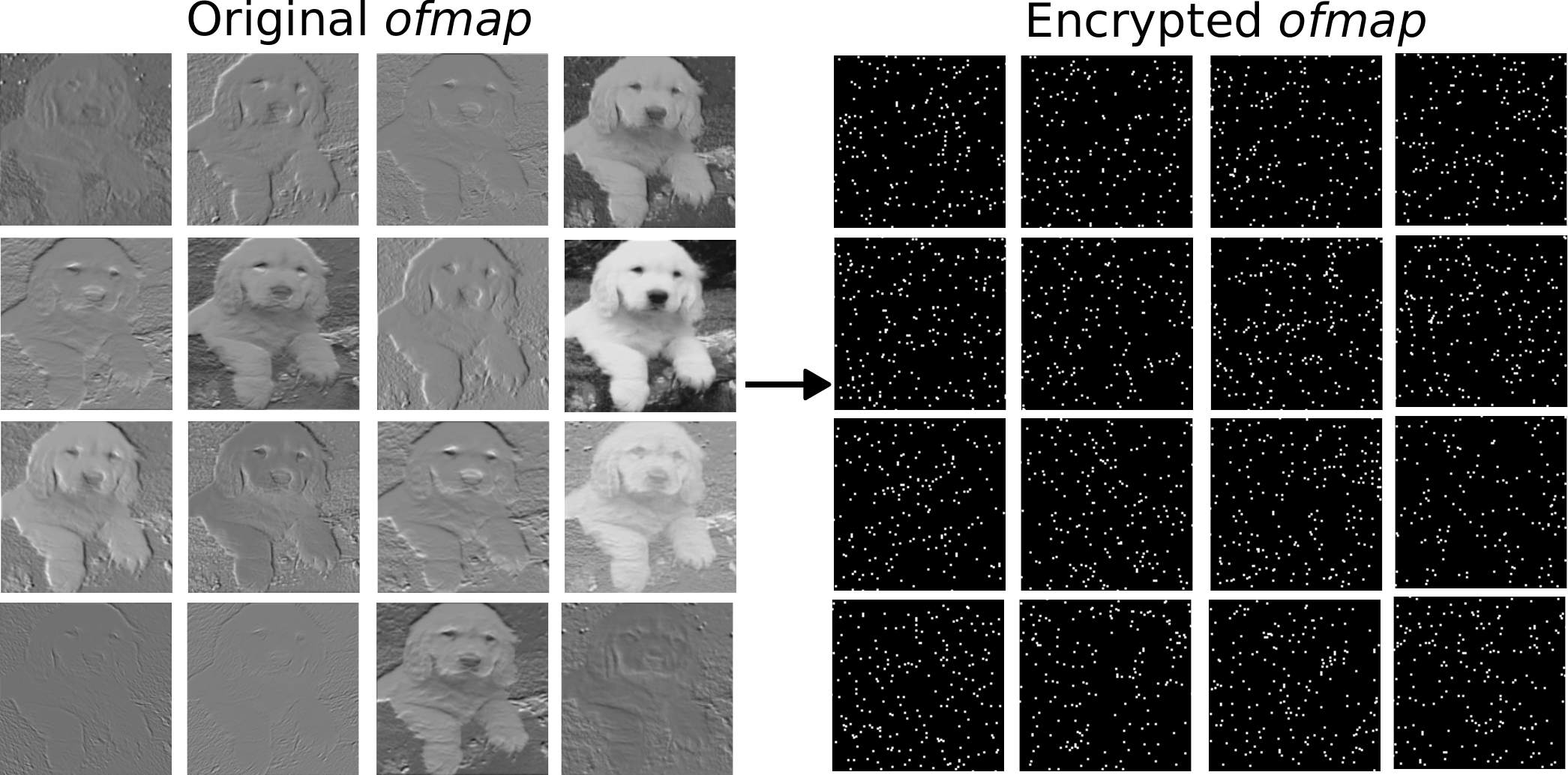}
    \caption{The intermediate \ofmaps of the {\em Vgg16} leak information regarding the original input}
    \label{fig:dog}
\vskip -4mm
\end{figure}

\vskip -3mm
The primary focus of many proposals lies in safeguarding the weights and the architecture of an
NN~\cite{securator,tnpu,guardnn}. It is essential to realize the significance of encrypting intermediate
\ofmaps as they can disclose information about the user inputs. We visualize the intermediate \ofmaps from the
second layer of {\em Vgg16} in Figure \ref{fig:dog}. Clearly, the \ofmaps leak information about the true input as we
can clearly observe.  In
order to overcome this, we encrypt the data using AES-CBC (part of the Securator component) and we observe that after encryption, 
correlations are not visible.
The additional security components for verifying the integrity and authenticity of the data provided
by {\em Securator} are also integrated into \fname. Given that the security guarantees of Securator~\cite{securator}
have been thoroughly validated in the original paper against Type C attacks, we shall assume the same. {\bf Note:
we use Securator as is and we don't make any change to it such that its security guarantees are diluted.}

\vskip -4mm

%\subsection{Limitations}
%The proposed scheme requires that the tiles should be compressible. It is highly unlikely that the weights and the \ofmaps are absolutely un-compressible. Several authors have established in their studies that the amount of sparsity in a NN model increases as we go deeper into the layers and the values of weights in a NN are generally very similar. Both of these facts will generate a high compression ratio.   

%% file: relatedWork.tex
\section{Related Work}
\label{sec:RW}

\subsection{Side-channel based DNN Architecture Extraction Attacks}  Table \ref{tab:relatedWork} show a summary of the most popular DNN architecture extraction
techniques.

\subsubsection{Dense NNs}
Hua et al.~\cite{reverse} propose a state-of-the-art SCA, which computes the potential dimension space of the model
architecture by observing memory access patterns and the memory traffic volume. They develop constraint
equations that relate such architectural hints to layer dimensions. The authors also consider the effects of dynamic
pruning to determine the filter weights. 
Cache Telepathy~\cite{cachetelepathy} is an identical attack that is based on the prime-probe
attack.DeepSniffer~\cite{deepsniffer}
and NeuroUnlock~\cite{neurounlock} are SCAs designed for GPUs that exploit the relationship between architectural hints
such as memory reads/write volumes and a DNN's architecture. Other than \cite{reverse}, the rest are not applicable to accelerators. Our scheme effectively stops the attack proposed in~\cite{reverse}.

\subsubsection{Sparse NNs} The authors~\cite{huffduff} exploit the boundary effect and the timing side-channel for
attacking sparse NN accelerators. The boundary effect occurs at the outermost edge of a convolutional layer, where the
portion of the filter outside the \ifmap does not contribute to the output layer. This effect causes a difference in the
NNZ elements between the \ofmap boundary and its center. This leaks information about the filter dimensions particularly.

\begin{table}[!htb]
    \centering
    \footnotesize
    \begin{tabular}{|p{17mm}|c|c|p{4mm}|p{17mm}|p{8mm}|}
    \hline
     \textbf{Attack}  & \multicolumn{2}{c|}{\textbf{Theft target}}  & \textbf{Sparse} & \textbf{Leaked} & \textbf{PF} \\
       \cline{2-3}
         & \textbf{Arch.} & \textbf{Params}   & \textbf{NN} & \textbf{metric} &\\
    \hline
    
     ReverseEngg \cite{reverse} & \fullcirc[1ex] &  \fullcirc[1ex] &  \emptycirc[1ex]  & MAP, MTV, Pruning info. & Acce.\\
     \hline
     CacheTelepathy \cite{cachetelepathy} & \fullcirc[1ex] &  \emptycirc[1ex]   & \emptycirc[1ex] &  Timing, MAP, MTV & GPP\\
      \hline
     
     DeepSniffer~\cite{deepsniffer} & \fullcirc[1ex] &  \emptycirc[1ex] &  \emptycirc[1ex] &  MAP, MTV, Timing & GPU\\
    \hline
    
     NeuroUnlock~\cite{neurounlock} & \fullcirc[1ex] &  \fullcirc[1ex] &  \emptycirc[1ex] &  MAP, MTV & GPU\\
    \hline
    
    HuffDuff~\cite{huffduff} & \fullcirc[1ex] &  \emptycirc[1ex]   & \fullcirc[1ex] &  Boundary effect, Timing & Sparse acce.\\
    \hline
    \end{tabular}
    \caption{Different Attacks on NNs. \fullcirc[1ex] shows that the property is satisfied, \emptycirc[1ex] shows it is not satisfied. \textbf{Sparse NN} represents whether the attack is applicable to sparse accelerators (\textbf{MAP}: Memory access patterns, \textbf{MTV}: Memory traffic volume).}
    \label{tab:relatedWork}
\end{table}

\subsection{DNN Architecture Protection}
Countermeasures (CMs) (see Figure~\ref{fig:CM}) can be categorized into three main categories: 

\circled{1} \textit{Using oblivious RAM (ORAM)} to obfuscate the access sequence. Despite the fact that several novel optimized designs for
ORAMs~\cite{oram1,oram2} have been proposed in the past, the use of ORAM for DNN applications makes them very slow. They incur
an additional overhead of $O(log N)$, which means that in order to access a single address, we must conduct at least
$log N$ accesses, where $N$ represents the total number of addresses~\cite{mitigating}. 

\circled{2} \textit{Using partial encryption} - Wang et al. propose
{\em NPUFort}~\cite{npufort}, a custom hardware accelerator that encrypts only security-critical features. The authors
claim that side-channel signatures are obscured due to the additional latency of the encryption operation. Sadly, this
scheme provides only partial security. 

\circled{3} \textit{Using additional dummy accesses} -  Liu et
al.~\cite{mitigating} propose to obfuscate access patterns by using both shuffling and additional dummy accesses. Similarly, {\em
NeurObfuscator}~\cite{neurobfuscator} also employs eight distinct obfuscation techniques such as layer deepening,
kernel widening, layer branching/skipping. There is a performance penalty and many side architectural hints such as the read-to-write
distance still remain intact.

\circled{4} \textit{Using address shuffling} - Liu et al.~\cite{mitigating} have
proposed CMs to shuffle the order of addresses. Their scheme still preserves read-to-write distances. Another issue is that the entire
shuffling requires a massive mapping table that cannot be stored on-chip~\cite{dnncloak}.  {\em
DNNCloak}~\cite{dnncloak} extends {\em NeurObfuscator} by obfuscating the dataflow between the accelerator and the DRAM
by deploying an additional address remapping function (lightweight Feistel cipher) alongside the layer obfuscation
circuit, which provides a random one-to-one mapping with minimal overheads. Shrivastava et al.~\cite{esl} have recently
proven that such shuffling-based CMs can be broken very quickly.

 \begin{figure}[H]
    \centering
    \includegraphics[scale=0.52]{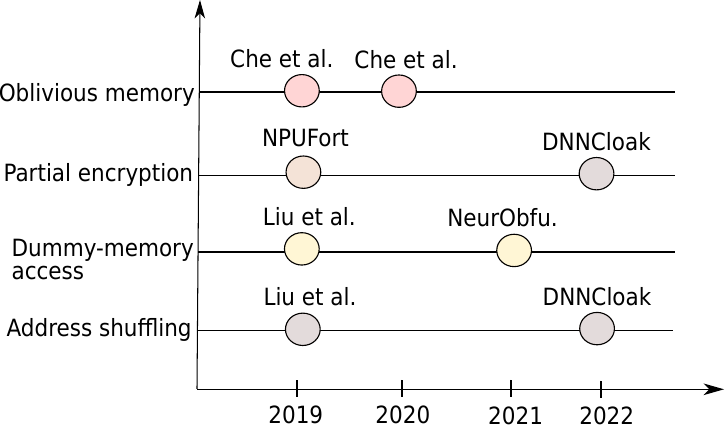}
    \caption{Different defences for model extraction attacks}
    \label{fig:CM}
\end{figure}

Other works on securing NN parameters include using secure custom accelerators ({\em TNPU}~\cite{tnpu}, {\em
GuardNN}~\cite{guardnn}, MGX~\cite{mgx} and Securator~\cite{securator}), offloading computations to a secure TEE ({\em
Slalom}~\cite{slalom}, {\em DarkNight}~\cite{darknight}, and {\em HybridTEE}~\cite{hybridtee}), and optimizing SGX
functionality for executing NN (Vessels~\cite{vessels}). These schemes are mainly for {\em Type C} attacks.

%DeepLaser~\cite{deeplaser} and DeepHammer~\cite{deephammer} attacks employs bit flips to compromise the integrity of a model. Similarly,

%% file: conclusion.tex
\section{Conclusion}
\label{sec:conclusion}
\vskip -3mm
This paper proposes a novel secure obfuscation technique that hides memory-access patterns to prevent recently proposed
advanced model stealing attacks. We utilize an intelligent combination of tiling, compression, encryption, and binning
to obfuscate architectural hints such as memory access patterns and NNZ patterns generated by repetitive
access patterns and boundary effects.  We
propose a novel compression scheme that serves the dual purposes of obfuscation and performance enhancement.  We
showed that the proposed scheme can protect sparse and dense NN models from Type A, B and C attacks.  In
addition, we also conducted a comprehensive information theoretic and statistical analysis to validate the security
guarantees of the proposed scheme. Our performance improvement over the nearest competitor (DNNCloak) was 47\%.